\documentclass[11pt]{article}
\pdfoutput=0 

\usepackage{jcappub} 
\usepackage{soul}
\usepackage[figtopcap]{subfigure}
\usepackage{amsmath,amssymb,dsfont}
\usepackage{hyperref}
\usepackage[T1]{fontenc} 
\newcommand\overcirc[1]{{\mathop{#1}\limits^{\circ}}}

\title{\boldmath Constant-roll Inflation in $f(T)$ Teleparallel Gravity}
\author[a,b]{A. Awad}
\author[c,d,1]{W. El Hanafy,\note{Corresponding author.}}
\author[c,d,e]{G.G.L. Nashed}
\author[f,g,h,i]{S.D. Odintsov}
\author[j,h,k]{V.K. Oikonomou}
\affiliation [a]{Department of Physics, School of Sciences and Engineering, American University in Cairo,
P.O. Box 74, AUC Avenue New Cairo, Cairo 11835, Egypt}
\affiliation[b]{Physics Department, Faculty of Science, Ain Shams University, Cairo 11566, Egypt}
\affiliation[c]{Centre for Theoretical Physics, The British University in Egypt, P.O. Box 43, El Sherouk City, Cairo 11837, Egypt}
\affiliation[d]{Egyptian Relativity Group (ERG), Cairo University, Giza 12613, Egypt}
\affiliation[e]{Mathematics Department, Faculty of Science, Ain Shams University, Cairo 11566, Egypt}
\affiliation[f]{ICREA, Passeig Luis Companys, 23, Barcelona 08010, Spain}
\affiliation[g]{Institute of Space Sciences (IEEC-CSIC), C. Can Magrans s/n, Barcelona 08193, Spain}
\affiliation[h]{Laboratory for Theoretical Cosmology, Tomsk State University of Control Systems and Radioelectronics (TUSUR),
Tomsk 634050, Russia}
\affiliation[i]{Institute of Physics, Kazan Federal University, Kazan 420008, Russia}
\affiliation[j]{Department of Physics, Aristotle University of Thessaloniki, Thessaloniki 54124, Greece}
\affiliation[k]{Tomsk State Pedagogical University, Tomsk 634061, Russia}
\emailAdd{adel.awad@bue.edu.eg}
\emailAdd{waleed.elhanafy@bue.edu.eg} \emailAdd{nashed@bue.edu.eg}
\emailAdd{odintsov@ieec.uab.es}
\emailAdd{v.k.oikonomou1979@gmail.com}
\abstract{We investigate in detail the implications of the
constant-roll condition on the inflationary era of a scalar field
coupled to a teleparallel $f(T)$ gravity. The resulting cosmological
equations constitute a reconstruction technique which enables us to
find either the $f(T)$ gravity which corresponds to a given
cosmological evolution, or the Hubble rate of the cosmological
evolution generated by a fixed $f(T)$ gravity. We also analyze in
some detail the phase space of the constant-roll teleparallel
gravity and we discuss the physical significance of the resulting
fixed points and trajectories. Also we calculate the observational
indices of a theory with given $f(T)$ gravity, and we discuss all
the implications of the constant-roll condition on these. As we
demonstrate, the resulting theory can be compatible with the current
observational data, for a wide range of values of the free
parameters of the theory.}
\arxivnumber{1710.00682}
\begin{document}
\maketitle
\flushbottom
\section{Introduction}\label{Sec:1}
Unquestionably, the inflationary era is of profound importance for
the description of the primordial cosmological evolution of our
Universe \cite{Starobinsky:1979ty,Starobinsky:1980te,Guth:1980zm,Mukhanov:1981xt,Linde:1981mu} (see also \cite{Lyth:1998xn,Linde:2007fr,Gorbunov:2011zzc}), and many
theoretical frameworks can successfully incorporate various version
of this early-time acceleration
\cite{Nojiri:2006ri,Capozziello:2011et,Capozziello:2010zz,Nojiri:2010wj,Clifton:2011jh,Nojiri:2017ncd}.
Most common is the scalar-tensor description, in which a scalar
field slowly rolls a nearly plateau like potential, however modified
gravity in various forms can describe an inflationary era. With
regards to the modified gravity description of inflation, it is also
possible to describe early and late-time acceleration within the
same theoretical framework \cite{Nojiri:2003ft,Bamba:2016gbu}.

One questionable feature of the scalar field description of
inflation, is that it is not possible to describe non-Gaussianities.
The existence of non-Gaussianities may be verified by future
observations of the primordial density perturbations. In some sense,
there are debatable arguments against the existence of
non-Gaussianities (see Ref. \cite{Chen:2010xka} for a review on
non-Gaussianities), that to our opinion are philosophically aligned
with the Occam's razor way of thinking, that is, the non-correlation
of the primordial modes is the simplest answer, and therefore
non-Gaussianities should be absent in the power spectrum. However,
this is a unilateral approach in the scientific problem at hand, and
a theory that aims to describe successfully the Universe should be
robust against any opposing future observation. In this context, it
was shown in Refs.
\cite{Inoue:2001zt,Tsamis:2003px,Kinney:2005vj,Namjoo:2012aa,Martin:2012pe,Motohashi:2014ppa,
Cai:2016ngx,Hirano:2016gmv,Cook:2015hma,Anguelova:2015dgt,Kumar:2015mfa,Odintsov:2017yud,
Odintsov:2017qpp,Nojiri:2017qvx,Fei:2017fub,Gao:2017owg,Oikonomou:2017xik}
that if the slow-roll condition is modified, non-Gaussianities can
be predicted even in the context of scalar-tensor theories of
inflation. Also in Refs.
\cite{Odintsov:2017yud,Odintsov:2017qpp,Oikonomou:2017xik} several
transition between constant and slow-roll eras were successfully
described. The implications of the constant-roll condition in $F(R)$
gravity were firstly studied in Ref. \cite{Nojiri:2017qvx}, and also
in a later publication \cite{Motohashi:2017vdc}, an alternative
approach was considered.

In this paper we shall investigate the implications of a
constant-roll inflationary era in the context of $f(T)$ teleparallel
gravity. The theoretical framework of $f(T)$ teleparallel gravity
has proved to be quite useful in cosmological and also astrophysical
applications, and for recent reviews we refer to
\cite{Cai:2015emx,Nojiri:2017ncd}. Particularly, late-time
acceleration in $f(T)$ gravity was studied in Refs.
\cite{Bamba:2013jqa,Bengochea:2008gz,Linder:2010py,Geng:2011aj,Otalora:2013tba,Chattopadhyay:2012eu,
Dent:2011zz,Yang:2010hw,Bamba:2010wb,Capozziello:2011hj,Geng:2011ka,Farajollahi:2011af,Cardone:2012xq,Bahamonde:2015zma},
and also inflationary and bouncing cosmology scenarios were studied
in
\cite{Cai:2011tc,Chen:2010va,Izumi:2012qj,Nashed:2014lva,Hanafy:2014bsa,Hanafy:2014ica,
Ferraro:2006jd,Bamba:2016gbu,ElHanafy:2017sih,Khurshudyan:2016qox}.
Also various astrophysical aspects of $f(T)$ gravity were addressed
in Refs.
\cite{Capozziello:2012zj,Paliathanasis:2014iva,Gonzalez:2011dr,Bohmer:2011si,Nashed:2013bfa,Ruggiero:2015oka,Nashed:2016tbj}
and in addition, the thermodynamics of $f(T)$ and other modified
gravities were studied in Ref. \cite{Bamba:2016aoo}. In view of the
various successful description of $f(T)$ gravity in both at a local
and global scales in the Universe, with this work we aim to
investigate thoroughly the implications of a constant-roll condition
in $f(T)$ gravity. We shall assume a scalar constant-roll condition
holds true, and we shall perform an in depth analysis of the various
implications on the $f(T)$ inflationary era. Particularly, we shall
demonstrate how the cosmological equations are altered in view of
the constant-roll condition, and we shall show that the resulting
formalism is actually a reconstruction mechanism that enables us to
either, fix the cosmological evolution and find the corresponding
$f(T)$ gravity which realizes the given cosmological evolution, or
to fix the $f(T)$ gravity and seek for the Hubble rate solution that
corresponds to this $f(T)$ gravity. In both cases, we shall assume
the existence of a scalar field which acts as the inflaton, and in
both cases we shall calculate the scalar potential that corresponds
to the constant-roll scenario under study. In the case that the
$f(T)$ gravity is fixed, we shall also calculate the cosmological
indexes corresponding to the power spectrum of the primordial
curvature perturbations, and particularly, the spectral index and
the scalar-to-tensor ratio, and accordingly we confront the results
with the latest Planck \cite{Ade:2015lrj} and BICEP2/Keck Array data
\cite{Array:2015xqh}. As we will show, the parameter that quantifies
the constant-roll era, enters to the final expressions of the
observational indices.

This paper is organized as follows: In section \ref{Sec:2} we
briefly review the essential features of $f(T)$ teleparallel gravity
and in section \ref{Sec:3} we present some characteristic results
from the minimally coupled scalar-$f(T)$ theory, that will be needed
in the sections to follow. In section \ref{Sec:4} we present the
reconstruction mechanism of constant-roll $f(T)$ gravity, and we
discuss how the constant-roll condition alters the formalism of
teleparallel gravity. In section \ref{Sec:5}, we fix the Hubble
evolution and we investigate which teleparallel gravity and which
potential can generate such an evolution. Also we perform a thorough
phase space analysis of the cosmological dynamical system,
discussing the physical meaning of the resulting fixed points. In
section \ref{Sec:6} we fix the functional form of the $f(T)$ gravity
and we investigate which Hubble evolution this generates, and also
we calculate in detail the observational indices of the
corresponding cosmological theory. Finally the results follow in the
end of this paper.
\section{Essential Features of Teleparallel Geometry}\label{Sec:2}
Before we get into the core of this paper, let us briefly present
some fundamental features of teleparallel geometry and gravity, for
details we refer the reader to the reviews
\cite{Cai:2015emx,Nojiri:2017ncd}. We consider a $4$-dimensional
smooth manifold $(M,\,h_{a})$, with  $h_{a}$ ($a=1,\cdots, 4$) being
four independent vector (tetrad) fields defined globally on $M$,
with the last condition actually being the realization of absolute
parallelism. The tetrad vector fields satisfy the tensor relation
$h_{a}{^{\mu}}h^{a}{_{\nu}}=\delta^{\mu}_{\nu}$ and also
$h_{a}{^{\mu}}h^{b}{_{\mu}}=\delta^{b}_{a}$, where ($\mu = 1,
\cdots, a$) are the coordinate components of the $a$-th vector field
$h_{a}$. By using the tetrad field, we can construct a
curvature-less (Weitzenb\"{o}ck) linear connection of the following
form $\Gamma^{\alpha}{_{\mu\nu}}\equiv
h_{a}{^{\alpha}}\partial_{\nu}h^{a}{_{\mu}}=-h^{a}{_{\mu}}\partial_{\nu}h_{a}{^{\alpha}}$.
Notably, the tetrad fields fulfill the teleparallel condition
$\nabla_{\nu}h_{a}{^{\mu}}\equiv 0$, where the operator
$\nabla_{\nu}$ is the covariant derivative with respect to the
Weitzenb\"{o}ck connection we defined above.

Also, the tetrad field can be used to construct the metric tensor on
the manifold $M$ by using $g_{\mu \nu} \equiv
\eta_{ab}h^{a}{_{\mu}}h^{b}{_{\nu}}$ with $\eta_{ab}$ being an
induced Minkowski metric on the tangent space of $M$. The inverse
metric is equal to $g^{\mu \nu} =
\eta^{ab}h_{a}{^{\mu}}h_{b}{^{\nu}}$ and subsequently the
Levi-Civita symmetric connection is
$\overcirc{\Gamma}{^{\alpha}}{_{\mu\nu}}= \frac{1}{2} g^{\alpha
\sigma}\left(\partial_{\nu}g_{\mu \sigma}+\partial_{\mu}g_{\nu
\sigma}-\partial_{\sigma}g_{\mu \nu}\right)$ can be defined, and in
effect, a Riemannian geometry can be defined. The torsion and the
contorsion tensors of the Weitzenb\"{o}ck connection are defined as
follows,
$T^\alpha{_{\mu\nu}}\equiv{\Gamma^\alpha}_{\nu\mu}-{\Gamma^\alpha}_{\mu\nu}={h_a}^\alpha\left(\partial_\mu{h^a}_\nu
-\partial_\nu{h^a}_\mu\right)$ and $K^{\alpha}{_{\mu\nu}} \equiv
\Gamma^{\alpha}_{~\mu\nu} -
\overcirc{\Gamma}{^{\alpha}}_{\mu\nu}=h_{a}{^{\alpha}}~
\overcirc{\nabla}_{\nu}h^{a}{_{\mu}}.$, where the covariant
derivative $\overcirc{\nabla}_{\nu}$ is defined with respect to the
Levi-Civita connection. The torsion tensor can be written in terms
of the contorsion tensor as $T_{\alpha \mu \nu}=K_{\alpha \mu
\nu}-K_{\alpha \nu \mu}$, while the inverse is equal to $K_{\alpha
\mu
\nu}=\frac{1}{2}\left(T_{\nu\alpha\mu}+T_{\alpha\mu\nu}-T_{\mu\alpha\nu}\right)$,
where $T_{\mu\nu\sigma} =
g_{\epsilon\mu}\,T^{\epsilon}_{~\nu\sigma}$\, and
\,$K_{\mu\nu\sigma} = g_{\epsilon\mu}\,K^{\epsilon}_{~\nu\sigma}$.

In teleparallel geometry, the torsion scalar is defined as follows,
$T=\frac{1}{4}T^{\alpha \mu \nu}T_{\alpha \mu
\nu}+\frac{1}{2}T^{\alpha \mu \nu}T_{\mu \alpha
\nu}-T^{\alpha}T_{\alpha}$, where $T^{\alpha}=T_{\rho}{^{\alpha
\rho}}$. The torsion scalar can be written in a compact form in the
following way,
\begin{equation}
T \equiv {T^\alpha}_{\mu \nu}{S_\alpha}^{\mu \nu},\label{Tor_sc}
\end{equation}
where the superpotential tensor ${S_\alpha}^{\mu\nu}$ is defined as
follows,
\begin{equation}
{S_\alpha}^{\mu\nu}=\frac{1}{2}\left({K^{\mu\nu}}_\alpha+\delta^\mu_\alpha{T^{\beta\nu}}_\beta-\delta^\nu_\alpha{T^{\beta
\mu}}_\beta\right),\label{superpotential}
\end{equation}
which is skew symmetric in the last pair of indices. The
teleparallel torsion scalar is equivalent to the Riemannian
curvature scalar $R$, up to a total derivative term. Consequently,
when $T$ is used in a Lagrangian instead of $R$ in Einstein-Hilbert
action, the resulting field equations are equivalent, and this is
actually the Teleparallel Equivalent of General Relativity (TEGR)
theory of gravity.
\section{Inflaton Minimally Coupled to $f(T)$ Teleparallel Gravity}\label{Sec:3}
In the context of $f(T)$ teleparallel gravity, the most successful
inflationary theories are those for which the inflaton $\phi$ is
minimally coupled to gravity, with the action being,
\begin{equation}\label{action}
    \mathcal{S}=\int d^{4}x |h|\left(\mathcal{L}_{g}+\mathcal{L}_{\phi}\right),
\end{equation}
where $|h|=\sqrt{-g}=\det\left({h}_\mu{^a}\right)$. The scalar field
part of the Lagrangian in Eq. (\ref{action}), namely
$\mathcal{L}_{\phi}$, is defined as follows,
\begin{equation}\label{scalar-Lag}
    \mathcal{L}_{\phi}=\frac{1}{2}\partial_{\mu}\phi~ \partial^{\mu}\phi-V(\phi),
\end{equation}
where $\partial^{\mu}=g^{\mu\nu}\partial_{\nu}$. By varying
$\mathcal{L}_{\phi}$ with respect to the metric, or equivalently
with respect to the tetrad fields \cite{deAndrade:1997ue}, enables
us to define the stress-energy tensor as follows,
\begin{equation}\label{scalar_Tmn}
 \mathop{\mathfrak{T}}\limits^{\phi}{_{\mu}}{^{\nu}}=h^{a}{_\mu}\left(-\frac{1}{h}\frac{\delta \mathcal{L}_{\phi}}{\delta h^{a}{_\nu}}\right)=\partial_{\mu}\phi~ \partial^{\nu}\phi-\delta_{\mu}^{\nu}
 \mathcal{L}_{\phi}\, ,
\end{equation}
which describes the matter content of the theory. We assume the
stress-energy tensor to have a perfect fluid form, so it can be
expressed as follows,
\begin{equation}\label{matter}
 \mathop{\mathfrak{T}}\limits^{\phi}{_{\mu\nu}}=\rho_{\phi} u_{\mu}u_{\nu}+p_{\phi}(u_{\mu}u_{\nu}+g_{\mu\nu}),
\end{equation}
where $u^{\mu}$ is the 4-velocity unit vector of the fluid. In most
cosmological applications, where a massless scalar field $\phi$ with
potential $V(\phi)$ is used, the unit vector is chosen to be normal
to spacelike hypersurfaces defined by $\phi =$ constant. In effect,
the stress-energy tensor (\ref{matter}) defines the scalar field
density $\rho_{\phi}$ and the corresponding pressure $p_{\phi}$ in
its rest frame, as follows,
\begin{equation}\label{scalar-dens-press}
    \rho_{\phi}=\frac{1}{2}\dot{\phi}^{2}+V(\phi), \quad p_{\phi}=\frac{1}{2}\dot{\phi}^{2}-V(\phi).
\end{equation}
In the above, we ignored an extra term which is generated by
existing anisotropies.

In the spirit of $f(R)$-gravity, in the context of which, one
replaces $R$ by an arbitrary function $f(R)$ in the Einstein-Hilbert
action, the TEGR has been generalized by replacing $T$ by an
arbitrary function $f(T)$
\cite{Bengochea:2008gz,Linder:2010py,Bamba:2010iw,Bamba:2010wb}. In
the natural units ($c=\hbar=k_{B}=1$), the $f(T)$ Lagrangian is
equal to,
\begin{equation}\label{gravity-Lag}
    \mathcal{L}_{g}=\frac{M_{p}^{2}}{2}\,f(T),
\end{equation}
where $M_{p}=2.4 \times 10^{18}$ GeV is the reduced Planck mass,
which can be related to the gravitational constant $G$ via
$M_{p}=1/\sqrt{8\pi G}\equiv 1/\kappa$. Upon varying the action
containing the Lagrangian  $\mathcal{L}_{g}$, with respect to the
tetrad fields, we obtain the tensor,
\begin{equation}\label{grav-Tmn}
    \tilde{H}_{\mu}{^\nu}=h^{a}{_\mu}\left(\frac{1}{h}\frac{\delta \mathcal{L}_{g}}{\delta h^{a}{_\nu}}\right)
    =\frac{M_{p}^2}{2} h^{a}{_\mu}\left(\frac{1}{h}\frac{\delta f(T)}{\delta h^{a}{_\nu}}\right)
\end{equation}
Upon rescaling, the tensor above takes the form
$H_{\mu}{^\nu}=\frac{1}{2}M_{p}^{-2} \tilde{H}_{\mu}{^\nu}$, and in
effect we have,
\begin{equation}\label{Hmn}
    H_{\mu}{^\nu}=S{_\mu}{^{\rho\nu}} \partial_\rho T f_{TT}+\left[\frac{h_{\mu}{^{a}}}{h} \partial_\rho \left( h S_a^{\verb| |\rho\nu} \right)- T_{\rho} S_\mu^{\verb| |\nu\rho}\right]f_{T}+\frac{1}{4} \delta^{\nu}_{\mu} f(T),
\end{equation}
where $f_{T}$ and $f_{TT}$, stand for $f_{T}=\frac{d f(T)}{d T}$ and
$f_{TT}=\frac{d^2 f(T)}{d T^2}$ respectively. By varying the action
(\ref{action}) with respect to the tetrad fields, using Eqs.
(\ref{grav-Tmn}) and (\ref{scalar_Tmn}), gives the following field
equations of $f(T)$ teleparallel gravity,
\begin{equation}\label{Field-equations}
    H_{\mu}{^\nu}=\frac{1}{2} M_{p}^{-2} \mathop{\mathfrak{T}}\limits^{\phi}{_{\mu}}{^{\nu}},
\end{equation}
or equivalently, by substituting from Eq. (\ref{Hmn}), we obtain,
\begin{equation}\label{field_eqns}
S_a^{\verb| |\rho\nu} \partial_\rho T f_{TT}+\left[\frac{1}{h} \partial_\rho \left( h S_a^{\verb| |\rho\nu} \right)-h_a^\lambda  T^\mu_{\verb| |\rho \lambda} S_\mu^{\verb| |\nu\rho}\right]f_{T}+\frac{1}{4} h_a^\nu f(T)=\frac{M_{p}^{-2}}{2} h_a^\mu \mathop{\mathfrak{T}}\limits^{\phi}{_{\mu}}{^{\nu}},
\end{equation}
It is clear that the general relativistic limit is recovered by
setting $f(T)=T$. We will assume that the background metric is a
flat Friedmann-Lema\^{\i}tre-Robertson-Walker (FLRW) metric, with
line element,
\begin{equation}\label{FRW-metric}
ds^2=dt^{2}-a(t)^{2}\delta_{ij} dx^{i} dx^{j},
\end{equation}
where $a(t)$ is the scale factor of the Universe. Thus, the vierbein
may take the following diagonal form,
\begin{equation}\label{tetrad}
{h_{\mu}}^{a}=\textmd{diag}\left(1,a(t),a(t),a(t)\right).
\end{equation}
This directly relates the teleparallel torsion scalar (\ref{Tor_sc})
to Hubble rate as follows,
\begin{equation}\label{TorHubble}
T=-6H^2,
\end{equation}
where $H\equiv \dot{a}/a$ is Hubble rate, and the ``dot'' denotes
differentiation with respect to the cosmic time $t$. Inserting the
vierbein (\ref{tetrad}) into the field equations (\ref{field_eqns})
for the scalar field matter fluid (\ref{matter}), the modified
Friedmann equations of the $f(T)$-gravity are,
\begin{eqnarray}
  \rho_{\phi} &=& \frac{M_{p}^{2}}{2}\left[f(T)+12 H^2 f_{T}\right], \label{FR1T}\\
  p_{\phi} &=& -\frac{M_{p}^{2}}{2}\left[f(T)+4(3H^2+\dot{H})f_{T}-48\dot{H}H^2 f_{TT}\right].\label{FR2T}
\end{eqnarray}
Independently from the above equations, one could choose an equation
of state to relate $\rho_{\phi}$ and $p_{\phi}$. Here, we choose the
simple barotropic case $p_{\phi}\equiv p_{\phi}(\rho_{\phi})=w_\phi \rho_\phi$.
Generally, any modified theory of gravity should be recognized as a
correction of the standard general relativistic gravity, so it is
convenient to transform from the matter frame, we have been using,
to the effective frame, which yields Einstein's gravity, in addition
to the higher order $f(T)$ teleparallel gravity. So we rewrite the
modified Friedmann equations in the case of $f(T)$-gravity, as
follows,
\begin{eqnarray}
{H}^2& =& \frac{M_{p}^{-2}}{3} \left( \rho_{\phi}+  \rho_{ T} \right)~\equiv \frac{M_{p}^{-2}}{3} \rho_{eff}, \label{MFR1}\\
2 \dot{{H}} + 3{H}^2&=& - M_{p}^{-2} \left(p_{\phi}+p_{ T }\right)\equiv -M_{p}^{-2} p_{eff}.\label{MFR2}
\end{eqnarray}
In this case, the density and pressure of the torsional counterpart
of $f(T)$ are defined by,
\begin{eqnarray}
  \rho_{T} &=& \frac{M_{p}^{2}}{2}\left(2Tf_T-T-f(T)\right), \label{Tor-density}\\
  p_{T} &=& \frac{M_{p}^{2}}{2}\frac{f(T)-Tf_T+2T^2f_{TT}}{f_{T}+2Tf_{TT}}.\label{Tor-press}
\end{eqnarray}
At the GR limit ($f(T)=T$), we have $\rho_{T}=0$ and $p_{T}=0$. In
the barotropic case, the torsion will have an equation of state,
\begin{equation}\label{torsion_EoS}
    w_{T}=-1+\frac{p_{T}}{\rho_{T}}=-1+\frac{(f(T)-2T f_{T})(f_{T}+2Tf_{TT}-1)}{(f(T)+T-2Tf_{T})(f_{T}+2Tf_{TT})}.
\end{equation}
To fulfill the conservation law, when the scalar field and the
torsion are minimally coupled, we have the following continuity
equations,
\begin{eqnarray}
  \dot{\rho}_{\phi}+3H(\rho_{\phi}+p_{\phi}) &=& 0, \label{sc-continuity}\\
  \dot{\rho}_{T}+3H(\rho_{T}+p_{T}) &=& 0. \label{tor-continuity}
\end{eqnarray}
Also the effective equation of state (EoS) parameter is defined as
follows,
\begin{equation}\label{eff_EoS}
w _{eff}\equiv \frac{p_{eff}}{\rho_{eff}}=-1-\frac{2}{3}\frac{\dot{H}}{H^2}.
\end{equation}
\section{Slow-roll and Constant-roll Inflation in $f(T)$ Teleparallel Gravity}\label{Sec:4}
In this section we shall investigate the qualitative and
quantitative consequences of a constant-roll inflationary era in
$f(T)$ teleparallel gravity.

It is a known fact that the modification of the Friedmann equations
due to $f(T)$ gravity, can be written as a one-dimensional autonomous
system of the form $\dot{H}\equiv \dot{H}(H)=\mathcal{F}(H)$, for a
general barotropic equation of state \cite{Awad:2017yod}. In this case, it is more
convenient to use the Hubble rate $H$ as an independent variable
instead of the torsion scalar $T$. Using Eq. (\ref{TorHubble}), the
modified Friedmann equations (\ref{FR1T}) and (\ref{FR2T}) can be
written as follows,
\begin{eqnarray}
  \rho_{\phi} &=& ~\frac{M_{p}^{2}}{2}\left(f-H f'\right), \label{FR1H}\\
  p_{\phi}    &=& -\frac{M_{p}^{2}}{2}\left(f-H f'-\frac{1}{3}\dot{H} f''\right)=\frac{M_{p}^{2}}{6}\dot{H} f''-\rho_{\phi}, \label{FR2H}
\end{eqnarray}
where $f\equiv f(H)$, $f'\equiv \frac{d f}{d H}$ and $f''\equiv
\frac{d^{2} f}{d H^{2}}$. Interestingly, Eq. (\ref{FR2H}) shows that
the Hubble parameter does not only decrease as in the GR limit, but
it can also increase without violating the weak energy condition (WEC),
$\rho_{\phi}+p_{\phi}\geq 0$. In particular, we have $\dot{H}>0$ with
$f''>0$, while $\dot{H}<0$ with $f''<0$. The last case includes the
particular value $f''=-12$ which produces the GR limit. By using Eq.
(\ref{scalar-dens-press}), we obtain the inflaton's kinetic term and
the scalar potential,
\begin{eqnarray}
  \rho_{\phi}&=&\frac{M_{p}^2}{2}\left(f-H f'\right) = \frac{\dot{\phi}^{2}}{2}+V(\phi), \label{sc-pot-term} \\
  \rho_\phi+p_\phi&=&\frac{M_{p}^{2}}{6}\dot{H} f''  \qquad     = \dot{\phi}^2. \label{sc-kin-term}
\end{eqnarray}
The last equation reads $(1+w_\phi)\rho_\phi=\frac{M_{p}^{2}}{6}\dot{H} f'' = \dot{\phi}^2$, where $w_\phi=p_\phi/\rho_\phi$, which shows that the kinetic term is non-negative as long as $w_\phi\geq -1$ and $\rho_\phi \geq 0$, that is in fact consistent with the WEC. Otherwise, if $w_\phi<-1$ and $\rho_\phi> 0$, the kinetic term will be negative which implies to a phantom case. In other words, the WEC in the case of an ordinary scalar field minimally coupled to $f(T)$ gravity can be written as $f-Hf'\geq 0$ and $\dot{H}f''\geq 0$.
Also, the continuity equation (\ref{sc-continuity}) is nothing but
the Klein-Gordon equation of motion for the inflaton in the FLRW
background,
\begin{equation}\label{Klein-Gordon}
    \ddot{\phi}+3H\dot{\phi}+\frac{\partial V}{\partial \phi}=0,
\end{equation}
which also can be obtained directly by varying the inflaton
Lagrangian $\mathcal{L}_{\phi}$,  appearing in Eq.
(\ref{scalar-Lag}), with respect to the scalar field $\phi$. In the
context of the slow-roll inflation approximation, two conditions
have been imposed to the dynamical variables, firstly the condition
which guarantees an accelerated expansion phase,
\begin{equation}\label{1st-slow-roll-cond}
    \frac{1}{2}\dot{\phi}^{2} \ll V(\phi),
\end{equation}
and secondly, the condition which makes the duration of the
inflationary era prolonged, which is,
\begin{equation}\label{2nd-slow-roll-cond}
    |\ddot{\phi}| \ll \left|\frac{\partial V}{\partial \phi}\right|.
\end{equation}
Substituting Eqs. (\ref{sc-kin-term}), (\ref{1st-slow-roll-cond})
and (\ref{2nd-slow-roll-cond}) into Eq. (\ref{Klein-Gordon}), it is
easy to show that the slow-roll potential in terms of the $f(T)$
gravity can be obtained as follows,
\begin{equation}\label{bdh}
V'\simeq -\frac{M_{p}^{2}}{2}H f''\, ,
\end{equation}
so by integrating we get,
\begin{equation}\label{slow-roll-pot-f(T)}
V(H) = V_{0}+\frac{M_{p}^{2}}{2}\left(f-Hf'\right).
\end{equation}
In the GR limit, that is when $f=-6H^2$, the above equation
reproduces the well-known relation
\begin{equation}\label{GR-SR-pot}
    V(\phi)\simeq V_{0}+3M_{p}^2H^{2}.
\end{equation}
It is useful to parameterize the inflationary Universe by defining
the first Hubble slow-roll index \cite{Nojiri:2017ncd} and its
running
\begin{equation}\label{Hubble-slow-roll}
    \epsilon_{1}=-\frac{d \ln H/dt}{H}, \quad \epsilon_{N+1}=\frac{d\ln \epsilon_{N}/dt}{H};\quad N \in \mathds{Z}^+.
\end{equation}
Since $H$ is almost constant during slow-roll inflationary era, the
first slow-roll  index is $\epsilon_{1}<1$. However, the slow-roll
inflation particularly requires also $\epsilon_{N}$ to be small,
that is, $|\epsilon_{N}| \ll 1$. In some other inflationary models,
the slow-roll conditions have been modified, and replaced by a
condition in which the term $|\ddot{\phi}|$ is no longer negligible.
This is known as the so-called ultra slow-roll condition
\cite{Tsamis:2003px},
\begin{equation}\label{ultra-slow-roll}
    \ddot{\phi}=-3H\dot{\phi}.
\end{equation}
This condition has been introduced in order to produce a potential
with an exact flat plateau (\ref{Klein-Gordon}). It has been shown
that the condition (\ref{ultra-slow-roll}) violates the slow-roll
approximation, where the running of the first Hubble slow-roll index
$|\epsilon_{2}|$ is no longer small \cite{Tsamis:2003px}. This makes
the ultra slow-roll models in fact to be some sort of a fast-roll
inflation model, like the ones of Ref. \cite{Motohashi:2014ppa}.
Interestingly enough, the ultra slow-roll inflation can produce a
scale invariant power spectrum \cite{Kinney:2005vj}. However, the
scalar (curvature) perturbations grow on the super-horizon energy
scale unlike the slow-roll inflation
\cite{Kinney:2005vj,Namjoo:2012aa} which disfavors the ultra
slow-roll condition \cite{Martin:2012pe}.

Along the research line of the ultra slow-roll inflation, the
constant-roll inflation scenario
\cite{Martin:2012pe,Motohashi:2014ppa,
Cai:2016ngx,Hirano:2016gmv,Cook:2015hma,Anguelova:2015dgt,Kumar:2015mfa,Odintsov:2017yud,
Odintsov:2017qpp,Nojiri:2017qvx,Fei:2017fub,Gao:2017owg,Oikonomou:2017xik}
is a modification of the slow-roll inflation scenario, in which case
the following condition holds true,
\begin{equation}\label{constant-roll}
    \ddot{\phi}=\beta H\dot{\phi},
\end{equation}
where the slow-roll condition can be recovered if $\beta\ll 1$,
while the ultra slow-roll is recovered by setting $\beta=-3$.
Remarkably, it has been shown that $\beta=-3/2$ is a critical value,
where the scalar perturbations grow for $\beta<-\frac{3}{2}$ and
decay for $\beta>-\frac{3}{2}$ at the super-horizon scale. Moreover,
the first Hubble slow-roll parameter satisfies $\epsilon \ll 1$
during the constant-roll inflation era, but its running satisfies
$|\epsilon_{N}|> 1$. Furthermore, in spite of using a single
inflaton in the constant-roll inflationary scenario, the local
non-Gaussianity consistency relation can be violated which makes the
constant-roll inflation phenomenologically distinguishable from the
slow-roll scenario.

The constant-roll inflation has been studied in the $f(R)$ gravity context in two different ways. In the first approach, the constant-roll condition (\ref{constant-roll}) is applied to the $f(R)$ modified Friedmann equations \cite{Nojiri:2017qvx}, and in this way one can obtain the $f(R)$ gravity which generates a constantly rolling scalar field, or the construction a constant-roll potential for a given $f(R)$ gravity. Also, in extended studies, the possible transition between slow-roll and constant-roll and also between constant-roll eras, has been investigated in Refs. \cite{Odintsov:2017yud,Odintsov:2017qpp,Oikonomou:2017xik}. In the second approach, the condition $\ddot{F}=\beta H_{J} \dot{F}$ is considered as some sort of generalization of the constant-roll condition, where $F=df/dR_{J}$ and $H_{J}$ and $R_{J}$ are Hubble and Ricci scalar in Jordan frame \cite{Motohashi:2017vdc}. In this paper, we investigate the condition (\ref{constant-roll}) within the framework of the $f(T)$ teleparallel gravity.
\subsection{The Solution $H(\phi)$ and the Scalar Potential $V(\phi)$ in Constant-roll $f(T)$
Gravity}\label{Sec:4.1}
In this section we shall present a fundamental technique that will
enable us to derive the function Hubble rate function $H(\phi)$ in
the context of constant-roll $f(T)$ teleparallel gravity. This will
also enable us to obtain the scalar potential $V(\phi)$.

Plugging $\dot{H}=\dot{\phi}\frac{dH}{d\phi}$ in Eq.
(\ref{sc-kin-term}), we obtain,
\begin{equation}\label{phidot}
    \dot{\phi}=\frac{M_{p}^{2}}{6}f'' H_{\phi},
\end{equation}
where $H_{\phi}=\frac{dH}{d\phi}$, $H_{\phi
\phi}=\frac{d^{2}H}{d\phi^{2}}$ and also we used the fact that the
second derivative of the inflaton field with respect to the cosmic
time is,
\begin{equation}\label{ddphi}
    \ddot{\phi}=\frac{M_{p}^{2}}{6}\left[f'' H_{\phi \phi}+f''' H_{\phi}^{2}\right]\dot{\phi}.
\end{equation}
Then by applying the constant-roll condition (\ref{constant-roll}),
we obtain the following differential equation,
\begin{equation}\label{generatingH}
    f'' H_{\phi \phi}+f''' H_{\phi}^{2}-\frac{6\beta}{M_{p}^{2}}H=0.
\end{equation}
The above equation represents a modified version of the original
work of the constant-roll inflation \cite{Martin:2012pe} due to the
contribution of the torsional counterpart of $f(T)$ gravity. Unlike the general relativistic version, here we have two unknown functions $f(H)$ and $H(\phi)$. So we can use (\ref{generatingH}) two ways: We may begin with some solution $H(\phi)$, then generate the $f(T)$ gravity which satisfies the constant-roll condition, see Sec. \ref{Sec:5}. Also, we may begin with some $f(T)$ theory, then generate the constant-roll Hubble $H(\phi)$, see Sec. \ref{Sec:6}. However, these two ways are not equivalent. By careful look to (\ref{generatingH}), one realizes that the differential equation is linear in $f(H)$ but it is non-linear in $H(\phi)$. For example, at the
GR limit, that is when, $f(H)=-6H^{2}$, the second term in the above differential
equation vanishes and then the equation reduces to the simple harmonic oscillator differential
equation obtained in \cite{Martin:2012pe}, where $H(\phi)$ is given as a linear
combination of exponential basis $e^{\pm\sqrt{-\beta/2}\phi}$. However, by substituting, conversely, the general relativistic solution $H(\phi)$ into (\ref{generatingH}), its second term cannot be made to vanish and one should not expect to generate only the GR theory. We will examine this case in more detail in Sec. \ref{Sec:5}.

Since the teleparallel torsion (Hubble) can be related directly to
the inflaton field, then we have $f'=f_{\phi}/H_{\phi}$,
$f''=\left(f_{\phi\phi}H_{\phi}-f_{\phi}H_{\phi\phi}\right)/H_{\phi}^{3}$,
and so on. In effect, the differential equation (\ref{generatingH})
can be rewritten as follows,
\begin{equation}\label{generatingH2}
   H_{\phi}^{2}~f_{\phi\phi\phi}-2H_{\phi}~H_{\phi\phi}~f_{\phi\phi}+\left(2H_{\phi\phi}^{2}-H_{\phi}~H_{\phi\phi\phi}\right) f_{\phi}
   =\frac{6}{M_{p}^{2}}H_{\phi}^{3}H,
\end{equation}
which is a non-homogeneous third order linear differential equation in $f(\phi)$. The general solution is given as the linear combination
\begin{equation}\label{f(phi)}
    f(\phi)=c_{3}+c_2 \int H_\phi d\phi +c_{1}\int \phi H_\phi d\phi + \frac{6\beta}{M_{p}^{2}} \int \int \int \left(H d\phi d\phi\right) H_{\phi}~ d\phi,
\end{equation}
where $c_{1}$, $c_{2}$ and $c_{3}$ are arbitrary constants. It is
clear that the $c_{2}$ term $\propto H$, acts as a divergence term
in the action (\ref{action}). Hereafter we omit this term, and also
for simplicity we take $c_{3}=0$. It worths to mention here that the last term in the above equation identifies the GR solution, while the $c_1$ term is linearly independent from the GR solution and not expected to be a trivial multiple of the GR solution. In another word, for a given general relativistic cosmic evolution $H(\phi)$, the above solution determines the corresponding constant-roll $f(T)$ gravity where the GR relativistic ($c_1=0$) and the general case ($c_1\neq 0$) are both allowed in modified gravity. In fact this case will be examined in Sec. \ref{Sec:5}.

Turning our focus to the scalar potential, by inserting
(\ref{phidot}) into (\ref{sc-pot-term}), the constant-roll potential
can be written as follows,
\begin{equation}\label{const-roll pot}
    V(\phi)=\frac{M_{p}^{2}}{2}\left[f-Hf'-\frac{M_{p}^{2}}{36}f''^{2} H_{\phi}^{2}\right].
\end{equation}
The above equation represents a modified version of the
constant-roll potential which has been previously obtained in for
example, in the context of a canonical scalar field gravity. This
can be shown clearly by taking $f(H)=-6H^{2}+F(H)$, in effect Eq.
(\ref{const-roll pot}) becomes,
\begin{equation}\label{const-roll pot2}
    V(\phi)=M_{p}^{2}\left[3H^{2}-2M_{p}^{2}H_{\phi}^{2}\right]+\underbrace{\frac{M_{p}^{2}}{2}\left[F-H F' -\frac{M_{p}^{2}}{36}(F''-24)F''H_{\phi}^{2}\right]}_{\textmd{$f(T)$ modification}}.
\end{equation}
In the above expression for the scalar potential, the second term on
the right hand side, is essentially the contribution of $f(T)$
gravity in the constant-roll potential. For $F(H)=0$, the
constant-roll potential takes the usual scalar tensor form appearing
in the related literature. We note that for any $f(T)$ gravity, the
constant-roll inflationary era can be quantified by making use of
Eqs. (\ref{generatingH}) and (\ref{const-roll pot}).

Alternatively, applying the constant-roll condition
(\ref{constant-roll}) to the Klein-Gordon equation
(\ref{Klein-Gordon}), we obtain,
\begin{equation}\label{Const-roll-KG}
    (3+\beta)H \dot{H} \phi'^{2}+V'=0,
\end{equation}
where $H$ here is an independent variable. Since
$\dot{\phi}=\dot{H}\phi'$, Eq. (\ref{sc-kin-term}) becomes
$\dot{H}\phi'^{2}=\frac{M_{p}^{2}}{6}f''$. Thus, the above
differential equation takes the form,
\begin{equation}\label{Const-roll-KG-f(T)}
    V'=-\frac{M_{p}^{2}}{6}(3+\beta)H f''.
\end{equation}
\begin{equation}\label{pot2}
    V(H)=V_{0}+\frac{M_{p}^{2}}{6}(3+\beta)\left[f-Hf'\right].
\end{equation}
The equation (\ref{pot2}) is an equivalent alternative to Eq.
(\ref{const-roll pot}) up to an additive constant term $V_0$ which can produced in (\ref{const-roll pot}) by including the constant term $c_3$ in the general solution (\ref{f(phi)}) or by assuming $f(T)\to f(T)-2\Lambda$ in the lagrangian (\ref{gravity-Lag}). However, Eq. (\ref{pot2}) can be used to reconstruct
constant-roll potentials directly for a given $f(T)$ gravity,
without knowing the generating function $H(\phi)$. Conversely, it
enables us to reconstruct the $f(T)$ gravity, which generates a
given constant-roll potential $V(H)$. We also note that the
constant-roll inflation can be fully determined by combining Eqs.
(\ref{generatingH}) and (\ref{Const-roll-KG-f(T)}). By comparing (\ref{const-roll pot}) and (\ref{pot2}), for $V_0=0$ case, we find that
$$ f''^2 H^2_\phi= -\frac{12}{M^2_p}\beta (f-H f').$$
The above equation can be used to constrain the viable range of the constant-roll parameter in $f(T)$ scenario. Since the left hand side is always non-negative and $f-Hf'\geq 0$ in the right hand side as required by the WEC, we identify the viable range of the constant-roll parameter as $\beta \leq 0$.

In the sections to follow, we investigate some particular
constant-roll inflationary models. Our investigation is two-fold:
first, we shall assume a particular constant-roll generating
function $H(\phi)$ which has been obtained in Ref.
\cite{Motohashi:2014ppa}. According to the $f(T)$ contribution in
(\ref{generatingH}), a modified constant-roll potential will be
obtained. This will eventually constrain the constant-roll parameter
$\beta$ when the observational indices are taken into account.
Second, we shall assume a particular form of $f(T)$ teleparallel
gravity, and we shall construct the corresponding constant-roll
potential. Also, we shall examine the compatibility of the model
obtained, with the current observational data.
\section{Reconstruction of $f(T)$ Gravity for Constant-roll Inflation}\label{Sec:5}
In this section, we shall specify the Hubble rate as a function of
the scalar field $H(\phi)$, and by using this and the reconstruction
techniques we presented in the previous sections, we shall find the
scalar potential and the $f(T)$ gravity that may generate $H(\phi)$.
We assume that $H(\phi)$ has the following form
\cite{Motohashi:2014ppa},
\begin{equation}\label{Mod1-Hphi}
    H(\phi)=M \cos \left(\sqrt{\frac{\beta}{2}}\frac{\phi}{M_{p}}\right),
\end{equation}
where $M$ is a scale characteristic of $H(\phi)$. In order to assure
the validity of the spacetime description of the model, we assume
that $M\leq M_{p}$, otherwise, quantum gravity effects should be
also taken into account. Substituting Eq. (\ref{Mod1-Hphi}) in Eq.
(\ref{generatingH2}), we get,
\begin{equation}\label{Mod1-f(phi)}
    f(\phi)=-3M^{2}\left[1+\cos\left(\sqrt{2\beta}\frac{\phi}{M_{p}}\right)\right]+c_{1}\left[2M_{p}\sin \left(\sqrt{\frac{\beta}{2}}\frac{\phi}{M_{p}}\right)-\sqrt{2\beta} \phi \cos \left(\sqrt{\frac{\beta}{2}}\frac{\phi}{M_{p}}\right)\right].
\end{equation}
The above equation will enable us to reconstruct the corresponding
constant-roll potential $V(\phi)$  and also, it can be used to
reconstruct the corresponding $f(T)$ theory that generates
$H(\phi)$. One can see that equations (\ref{Mod1-Hphi}) and (\ref{Mod1-f(phi)}) satisfy the $f(T)$ constant-roll condition (\ref{generatingH}). However, the $f'''$-term in (\ref{generatingH}) cannot be made to vanish by taking the general relativistic solution (\ref{Mod1-Hphi}), so one should expect a solution different from the GR theory. As a matter of fact, the general solution (\ref{Mod1-f(phi)}) reduces to the GR theory for $c_1=0$. However, the non-zero values of $c_1$ are allowed too, and consequently a new modified constant-roll potential can be generated in the $f(T)$ gravity scenario.

\begin{figure}
\centering
\subfigure[~$\beta<0$]{\label{fig:Mod1-pot1}\includegraphics[scale=.38]{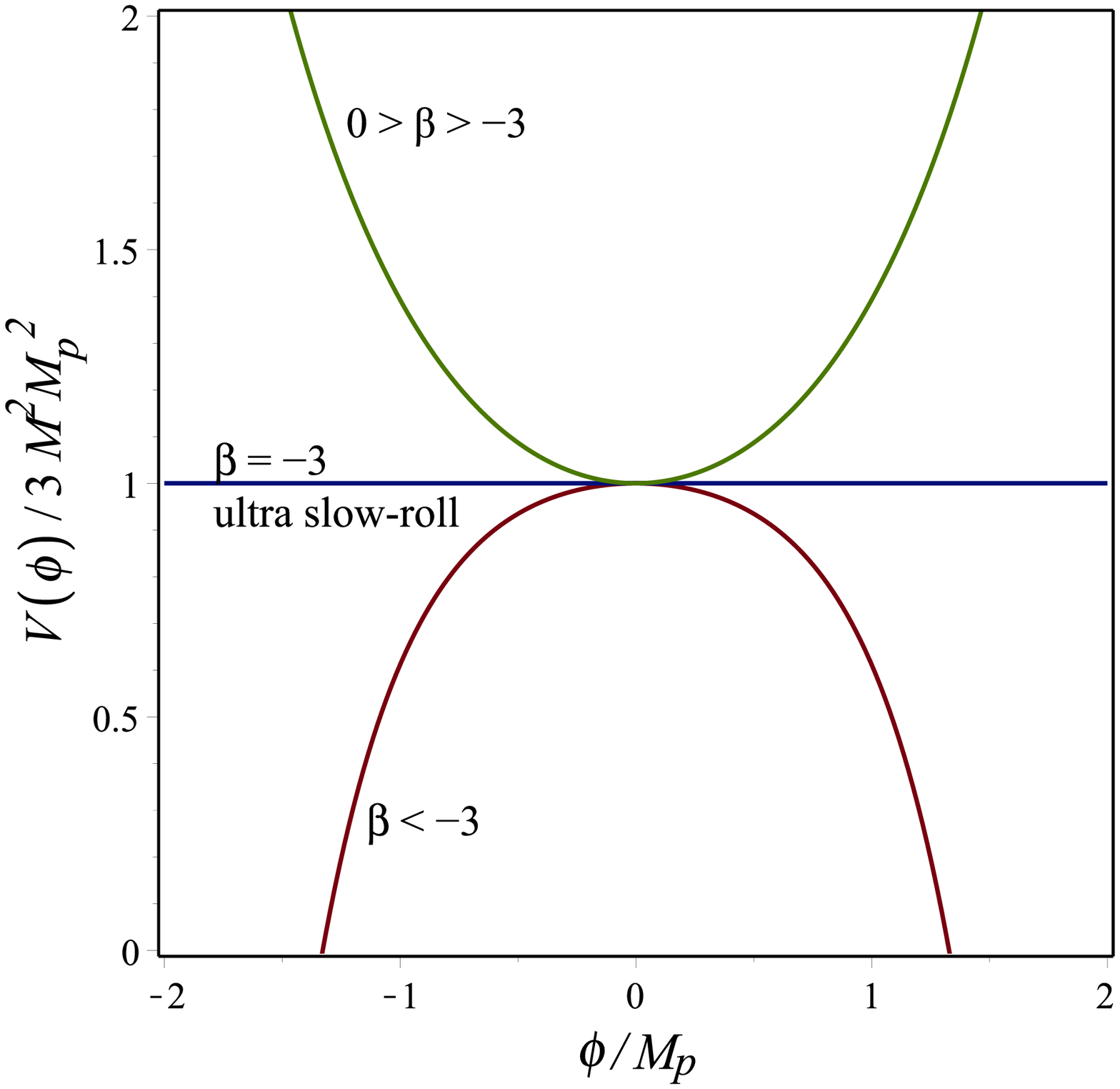}}\hspace{.2cm}
\subfigure[~$\beta>0$]{\label{fig:Mod1-pot2}\includegraphics[scale=.38]{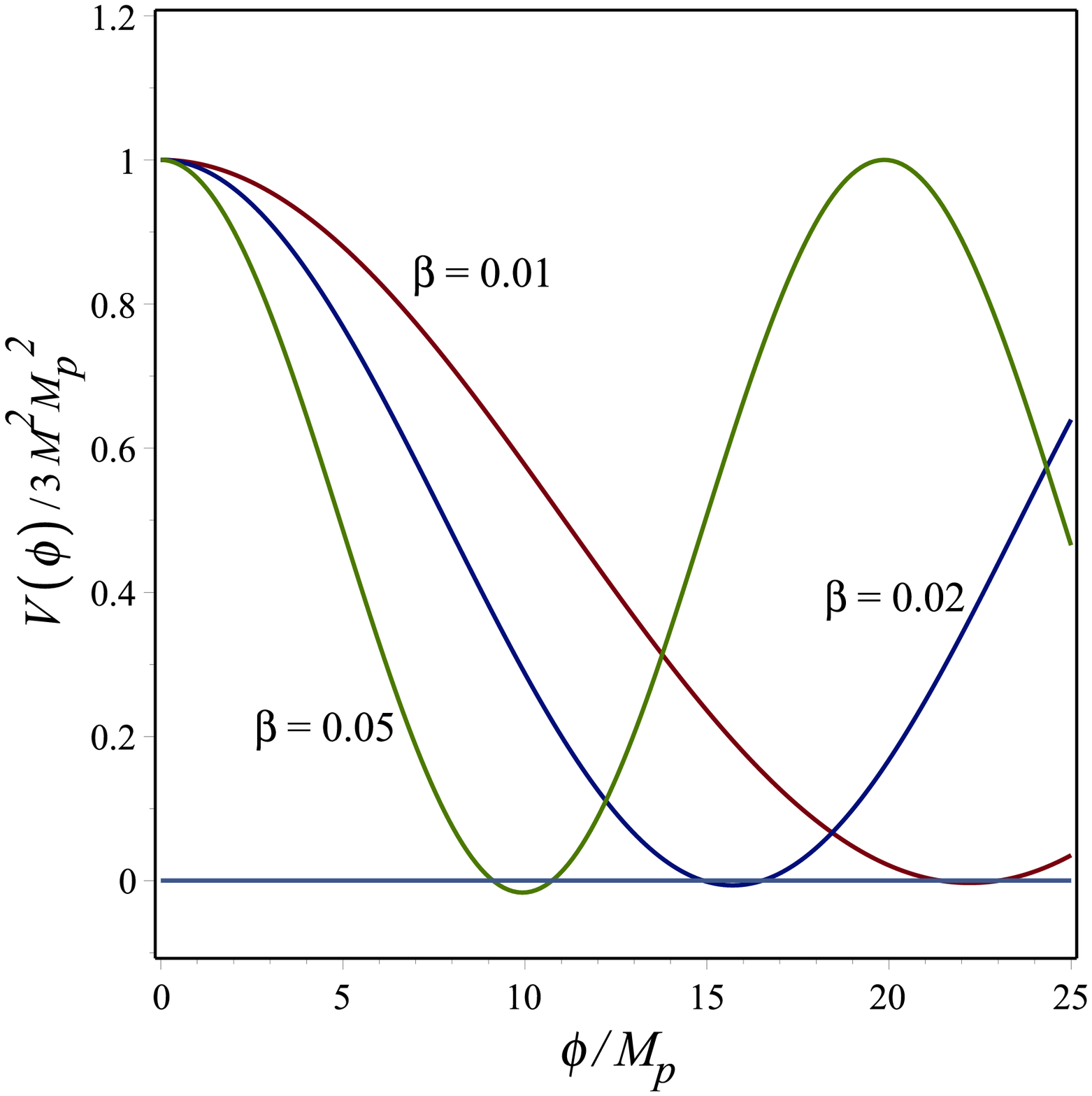}}
\caption[figtopcap]{\textit{Possible potential patterns of Eq.} (\ref{Mod1-Vphi}):
\subref{fig:Mod1-pot1}\textit{ For $\beta<-3$ the potential has no attractor. For $\beta=-3$, the constant potential of the ultra slow roll model is achieved. For $0<\beta<-3$ the potential has an attractor. We note that the present potential patterns are qualitatively similar to that has been previously obtained in Ref. \cite{Motohashi:2014ppa} regardless the value of the constant $c_1$};
\subref{fig:Mod1-pot2} \textit{For $\beta>0$, the cosine potential pattern is achieved where $\beta$ determines its frequency. We note that we keep our discussion in this case limited to small values of $c_1$, which match the physical scenarios as we will explain latter on when we discuss the phase portraits of the model.}}
\label{Fig:Mod1-potentials}
\end{figure}

By substituting Eqs. (\ref{Mod1-Hphi}) and (\ref{Mod1-f(phi)}) in
Eq. (\ref{const-roll pot}), we obtain the scalar potential,
\begin{eqnarray}\label{Mod1-Vphi}
\nonumber    V(\phi)&=&3 M^{2} M_{p}^{2}\left[1-\frac{3+\beta}{6}\left\{1-\cos \left(\sqrt{2\beta}\frac{\phi}{M_{p}}\right)\right\}\right]\\
&-&\frac{c_{1}M_{p}^{2}}{36}\left[c_{1}\beta \frac{M_{p}^{2}}{M^{2}}-12M_{p}(3+\beta) \sin \left(\sqrt{\frac{\beta}{2}}\frac{\phi}{M_{p}}\right)\right].
\end{eqnarray}
The quantity in the second line of Eq. (\ref{Mod1-Vphi}) corresponds
to the $f(T)$ contribution to the cosmic evolution
(\ref{Mod1-Hphi}). For $c_1=0$, the potential reduces to the one
which has been obtained in Ref. \cite{Motohashi:2014ppa}. As a matter of fact, in the $\beta<0$ case, which is represented by Fig. \ref{Fig:Mod1-potentials}\subref{fig:Mod1-pot1}, the qualitative potential patterns of $c_1\neq 0$ are similar to that have been obtained in Ref. \cite{Motohashi:2014ppa}, when $c_1=0$. For $\beta>0$, the cosine potential patterns, when $c_1$ has small values, see Fig. \ref{Fig:Mod1-potentials}\subref{fig:Mod1-pot2}, are similar to that have been obtained in Ref. \cite{Motohashi:2014ppa} when $c_1=0$. These case matche the cosine natural inflation model \cite{Freese:2014nla} with a negative cosmological constant. On the contrary, for large values of $c_1$ different patterns can be obtained. We note that we keep our discussion in the present work limited to small values of $c_1$, which match the physical scenarios as we will explain latter on, in Sec. \ref{Sec:5.1}, when we discuss the phase portraits of the model.

In the general case ($c_1\neq 0$), one can examine the consistency of the new solution. In order to do that, we use the chain rule $f''=\left(f_{\phi\phi}H_{\phi}-f_{\phi}H_{\phi\phi}\right)/H_{\phi}^{3}$ and (\ref{phidot}), then write
$$\dot{\phi}=\frac{\sqrt{2\beta}M_p}{6M}\left[-c_1 M_p+6M^2 \sin \left(\sqrt{\frac{\beta}{2}}\frac{\phi}{M_p}\right)\right].$$
However, by making use of the above equation in addition to (\ref{Mod1-Hphi}), (\ref{Mod1-f(phi)}) and (\ref{Mod1-Vphi}), we verify the consistency of the obtained solution with the constant-roll Klein-Gordon equation $(3+\beta)H\dot{\phi}+V_\phi=0$.\\

In order to explore the physical meaning of the constant-roll parameter $\beta$, we use Eqs. (\ref{sc-pot-term}) and (\ref{sc-kin-term}) to obtain the following differential equation,
\begin{equation}\label{Mod1-phase-portrait1}
    \dot{H}=6~\frac{f-H f'-2 V(\phi)/M_p^2}{f''}.
\end{equation}
For the case $V_0=0$, we substitute the constant-roll potential in the $f(T)$ gravity (\ref{pot2}) into (\ref{Mod1-phase-portrait1}). Thus, we write
\begin{equation}\label{Mod1-phase-portrait2}
    \dot{H}=-2\beta~\frac{f-Hf'}{f''}=\mathcal{F}(H).
\end{equation}
In fact, the above equation represents a one-dimensional autonomous
system, since $\dot{H}$ can be written explicitly in terms of $H$.
The above relation can clearly dictate the role of the constant-roll
parameter $\beta$. As it can be easily shown, the parameter $\beta$
is strongly related to the inflaton equation of state. By assuming a
linear barotropic equation of state, that is, of the form
$p_{\phi}=w_{\phi}\rho_{\phi}$, by combining Eqs.
(\ref{sc-pot-term}), (\ref{sc-kin-term}) and
(\ref{Mod1-phase-portrait2}), we obtain,
\begin{equation}\label{Mod1-beta}
    \beta=-\frac{3}{2}(1+w_{\phi}).
\end{equation}
In conclusion, For $V_0=0$ case, the constant-roll parameter $\beta$ fixes the EoS parameter $w_\phi$ of the inflaton. By using Eq. (\ref{Mod1-beta}), we  classified all the different
inflationary scenarios in Table \ref{Table:beta-inflations}.\\
\begin{table}[h]
  \centering
  \caption{Classifications of possible inflationary scenarios according to Eq. (\ref{Mod1-beta}).}\label{Table:beta-inflations}
\begin{tabular}{cccc}
\hline\hline
 EoS & $\beta$ & Model & Curvature perturbations  \\
\hline
  $w_{\phi}>1$   & $\beta<-3$      & Non-attractor   & growing \\
  $w_{\phi}=1$   & $\beta=-3$      & ultra slow-roll & growing \\
  $0<w_{\phi}<1$ & $-3<\beta<-3/2$ & attractor & growing \\
  $w_{\phi}=0$   & $\beta=-3/2$    & Cosh potential & growing \\
  $-1<w_{\phi}<0$& $-3/2<\beta<0$  & attractor & decaying \\
  $w_{\phi}=-1$  & $\beta=0$       & slow-roll & decaying \\
  $w_{\phi}<-1$  & $\beta>0$       & Cosine potential & decaying \\
\hline\hline
\end{tabular}
\end{table}

Interestingly enough, the constant-roll inflationary scenario
becomes identical to the slow-roll, when $w_{\phi}=-1$, in which
case $\beta=0$. Also, when the constant-roll parameter $\beta$ takes
positive values, then the inflaton has a phantom EoS $w_{\phi}<-1$.
It is worth mentioning that if the $V_0\neq 0$ case has been considered in Eq. (\ref{pot2}), it will modify the phase portrait (\ref{Mod1-phase-portrait2}). In this case, we have
\begin{equation}
\nonumber    \dot{H}=-2\beta~\frac{f-Hf'}{f''}-12\frac{V_0}{M_p^2 f''}=\mathcal{F}(H).
\end{equation}
Following the same procedure above, we relate the constant-roll parameter and the EoS parameter $w_\phi$ by
$$w_\phi(H)=-\left(1+\frac{2}{3}\beta\right)-\frac{4V_0}{M_p^2(f-Hf')}.$$
In this case, the inflaton will have a dynamically varying EoS, where the Universe can interpolate between different scenarios. This is an
indirect approach to the transition problems studied in Refs.
\cite{Odintsov:2017yud,Odintsov:2017qpp,Oikonomou:2017xik}. Remarkably, in the latter case, i.e. $V_0\neq 0$, the constant-roll parameter $\beta$ may have positive values, while the WEC is still functioning successfully, see \cite{Motohashi:2014ppa,Motohashi:2017aob,Motohashi:2017vdc}.

\begin{figure}[hb]
\centering
\subfigure[~$c_{1}<0$]{\label{fig:unified-history}\includegraphics[scale=.35]{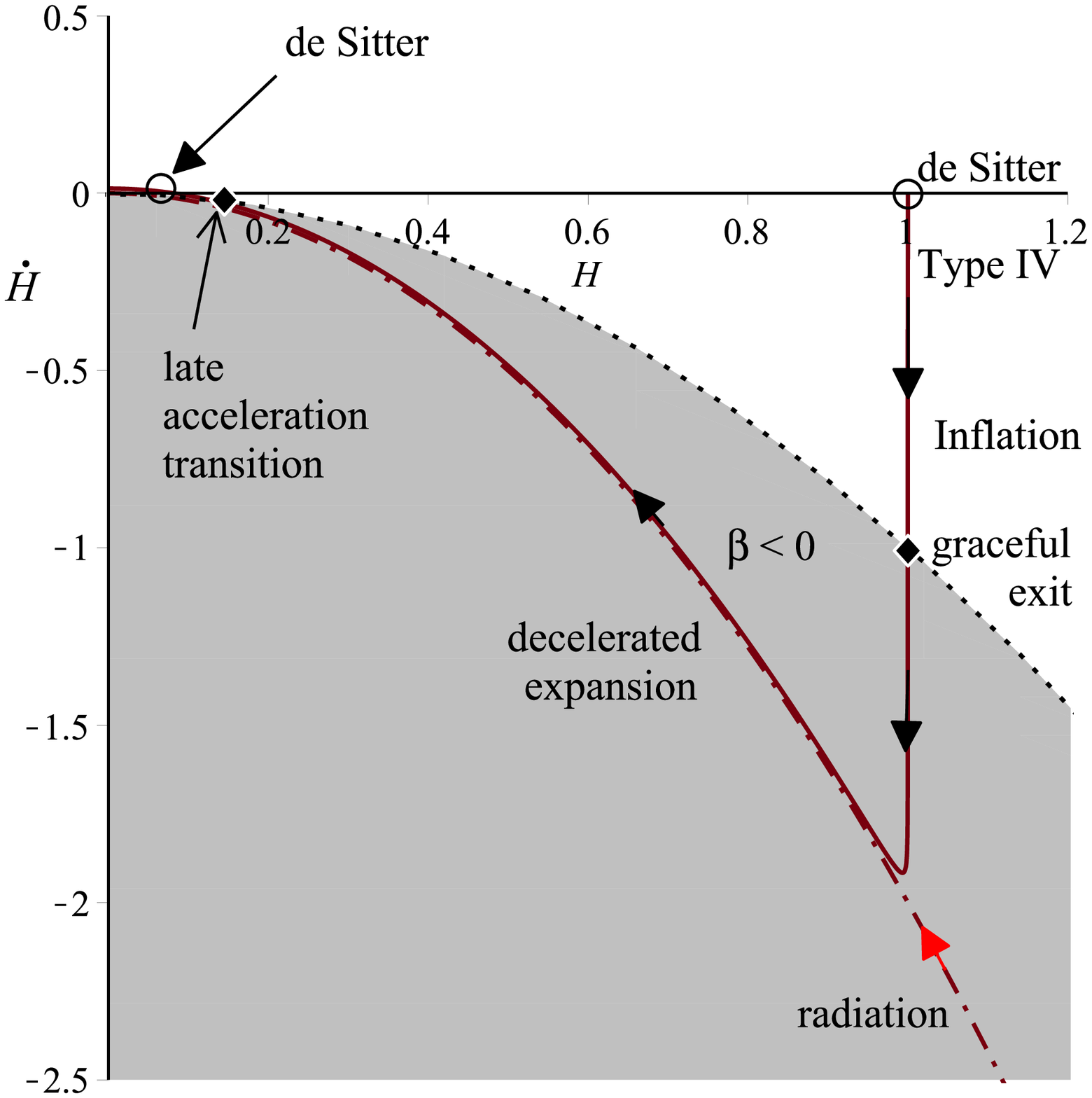}}\hspace{1.5cm}
\subfigure[~$c_{1}>0$]{\label{fig:cyclic}\includegraphics[scale=.35]{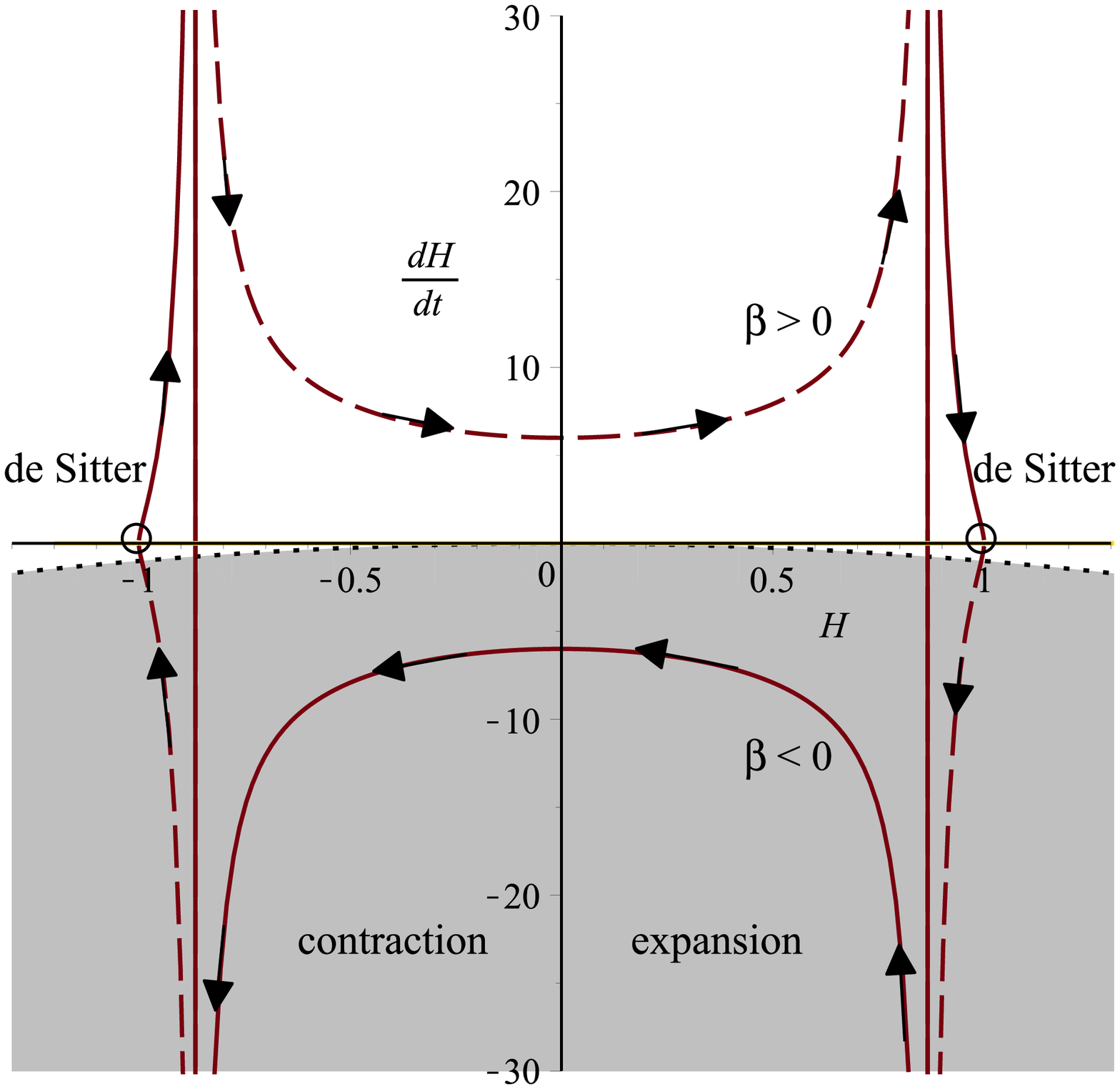}}
\caption[figtopcap]{\textit{Schematic plots of possible cosmic evolutions of the phase portrait} (\ref{Mod1-phase-portrait3});
\subref{fig:unified-history}\textit{ For $c_{1}<0$, $H>0$, and $\beta<0$: The universe interpolates between two de Sitter (fixed points) spaces, instead of big bang initial singularity there is type IV de Sitter state};
\subref{fig:cyclic} \textit{For $c_{1}>0$ and $\beta<0$: The universe interpolates between two sudden (type II) singularities where junction conditions are applicable in this model, which leads to a cyclic universe. We take $M=1$ and $M_p=1$.}}
\label{Fig:phase-portrait}
\end{figure}

For the model at hand, by using the inverse of Eq.
(\ref{Mod1-Hphi}), which yields $\phi \equiv \phi(H)$, we can
rewrite Eq. (\ref{Mod1-f(phi)}) in terms of $H$ as follows,
\begin{equation}\label{Mod1-f(H)}
    f(H)=-6H^{2}+\frac{2c_{1}M_{p}}{M}\left[\sqrt{M^2-H^2}-2H \arccos \left(\frac{H}{M}\right) \right].
\end{equation}
Thus, the constant-roll $f(T)$ gravity imposes the constraint $|H|
\leq M \leq M_{p}$ on the Hubble parameter. This constraint ensures
the physical consistency of the theory, since the Hubble rate cannot
exceed the Planck mass $M_p$. The first term in Eq.
(\ref{Mod1-f(H)}) reproduces the GR limit, while the term
proportional to  $c_{1}$, corresponds to the $f(T)$ gravity
modification. In effect, the differential equation of Eq.
(\ref{Mod1-phase-portrait2}) reads,
\begin{equation}\label{Mod1-phase-portrait3}
    \dot{H}=-2\beta \frac{\left(c_{1}M_{p}\sqrt{M^{2}-H^{2}}+3 M H^2\right)\sqrt{M^{2}-H^{2}}}{c_{1}M_{p}-6M\sqrt{M^{2}-H^{2}}}.
\end{equation}
Obviously, the choice $\beta=0$ implies that $\dot{H}=0$, which
matches exactly the de Sitter solution of slow-roll inflation. On
the other hand, the choice $c_{1}=0$ implies that $\dot{H}=\beta
H^{2}$, which matches exactly the standard cosmology where $\beta <0$. In this sense,
the non-null values of $c_{1}$ and $\beta$ could provide a cosmic
evolution interpolating between these two cases. In Fig.
\ref{Fig:phase-portrait}, we plot different phase portraits
corresponding to solutions of Eq. (\ref{Mod1-phase-portrait3}), for
various choices of the parameters $c_{1}\neq 0$ and $\beta\neq 0$.
%
\subsection{Analysis of the Phase Space}\label{Sec:5.1}
Let us now study in some detail the phase space of the cosmological
dynamical system of  Eq. (\ref{Mod1-phase-portrait3}). For $c_{1}<0$
and $\beta<0$, and during the era for which  $H>0$, the dynamical
system has two fixed points at,
$$H_{upper-fix}=M, \quad H_{lower-fix}= \frac{\sqrt{-2c_{1}M_{p}}}{6M}\sqrt{c_{1}M_{p}+\sqrt{c_{1}^{2}M_{p}^{2}+36 M^{4}}},$$
Obviously, the choice $c_{1}=0$ shifts the lower fixed point to be
that of a Minkowski Universe $H=0$, just as in the GR limit.
However, the choice $c_{1}<0$, enforces the Universe to evolve
towards the de Sitter solution $H > 0$ instead of the Minkowski.
Actually, the modification that the $f(T)$ gravity introduces to the
cosmological equations, namely (\ref{Mod1-f(H)}), makes easy to
interpret the de Sitter solution at the small $H$ regime as
late-time acceleration. In Fig.
\ref{Fig:phase-portrait}\subref{fig:unified-history}, the plot shows
that the Universe interpolates between two de Sitter phases, as the
Hubble rate decreases.  In the standard model of cosmology, the
Universe begins with an initial crushing-type singularity where the
Hubble rate and its derivative blow-up, that is, $H\to \infty$ and
$\dot{H}\to \infty$ at $t=0$ finite time. Interestingly enough, our
model imposes an upper bound for the Hubble parameter to be
$H_{max}=M$, which can be chosen to be consistent with a maximum
energy density at the Planck scale. Although, this maximum value of
Hubble rate is at a fixed point, this can be reached at a finite
time, which is,
\begin{equation}\label{Mod1-finite-time}
    t_{i}=\int_{H_{i}}^{M} \frac{dH}{\dot{H}}=finite, \quad H_{lower-fix}<H_{i}<M.
\end{equation}
In fact, at the upper fixed point the slope of the phase portrait $\frac{d\dot{H}}{dH}$ diverges which indicates the presence of a finite-time singularity of Type IV at that point \cite{Awad:2017yod} (see also \cite{Awad2013,ElHanafy:2017xsm,ElHanafy:2017sih,Hohmann:2017jao}). Alternatively, from Eq. (\ref{Mod1-phase-portrait3}), it can be shown that at the upper fixed point $H=M=finite$, $\dot{H}=0=finite$ but $\ddot{H}$ diverges. In this case, we call this fixed point a de Sitter of Type IV. In conclusion, the model replaces the initial big bang singularity with Type IV de Sitter singularity. As it is clear from Fig.
\ref{Fig:phase-portrait}\subref{fig:unified-history}, at the large
Hubble rate regime, the Universe undergoes an accelerating expansion
(unshaded region) at early-time, and also it exits into a FLRW
deceleration era (shaded region). The Hubble rate value at the
graceful exit time instance, can be identified as the cutting point
of the phase portrait with the zero acceleration curve, that is,
when $\dot{H}=-H^{2}$. For $\beta=-2$, the phase portrait matches
exactly the radiation phase portrait of standard Big Bang cosmology.
This feature is important in order to ensure a successful thermal
history. In addition, at the small Hubble rate regime, the phase
portrait cuts the zero acceleration curve once more towards a future
de Sitter fixed point $H_{lower-fix}$, and this fixed point is reached
asymptotically. In effect the Universe is free from any future
finite time singularities. At the second cutting point, the Hubble
rate value can be chosen as $H_{tr}\sim 100-120$ km/s/Mpc, so at a
redshift $z_{tr}\sim 0.6-0.8$, in order to be comparable with the
$\Lambda$CDM model at late times, which subsequently constrains $c_1$ to be small.

In summary, our model can provide a unified cosmic history of the
early and late-time acceleration eras. Also, the FLRW decelerated
phase is compatible with the standard cosmology. Moreover, the
late-time acceleration, compatible with $\Lambda$CDM model, is
realized without using a cosmological constant.

Now let us discuss another interesting solution of our dynamical
system, and for $c_{1}>0$, the fixed points of
(\ref{Mod1-phase-portrait3}) are $H_{fix}=\pm M$. The phase
portraits of the $c_{1}>0$ case are given in Fig.
\ref{Fig:phase-portrait}\subref{fig:cyclic}. We consider the more
physically appealing scenario $\beta<0$. From (\ref{Mod1-phase-portrait3}), we find that $\dot{H}$ diverges at finite Hubble values
$$H_{s\pm}=\pm \frac{\sqrt{36M^{4}-c_{1}^{2}M_{p}^{2}}}{6M}.$$
We conclude that these singularities are of type II (sudden singularities). Interestingly, the geodesics at these singular points are well behaved,
whereas the first derivative of the scale factor is finite and subsequently the Christoffel symbols are regular. For small nonnegative $c_1\ll 1$, we have $H_s\sim \pm M$ which near the maximum Hubble value. In Fig. \ref{Fig:phase-portrait}\subref{fig:cyclic}, we exaggerate the gap between de Sitter points and the sudden singularities in order to get clear view. The phase portrait of $c_1>0$ is similar to that have been obtained in Ref. \cite{Awad:2015syb,Awad:2017sau}, where junction conditions have been used to joint two branches of the solution. In this case, the universe will be oscillating between the two sudden singularities $H_{s\pm}$ with not attractors. However, the consistency of the extension of spacetime through joining the two branches of solutions with the junction conditions and the geodesic extension need to be checked. We note here that both scenarios, i.e. $c_1<0$ and $c_1>0$, are expected to be viable when $c_1$ has small values. This justifies the realization of the potential patterns in Fig. \ref{Fig:Mod1-potentials}.\\

Now let us turn our focus on the $f(T)$ gravity that generates the
cosmic evolution  (\ref{Mod1-Hphi}), so by substituting Eq.
(\ref{TorHubble}) in (\ref{Mod1-f(H)}), we obtain,
\begin{equation}\label{Mod1-f(T)}
    f(T)=T+\sqrt{\frac{2}{3}}\frac{c_{1}M_{p}}{M}\left[\sqrt{6M^2+T}-\sqrt{-6T} \arccos \left(\frac{\sqrt{-6T}}{6M}\right)\right].
\end{equation}
\begin{figure}[h]
\begin{center}
\includegraphics[scale=0.35]{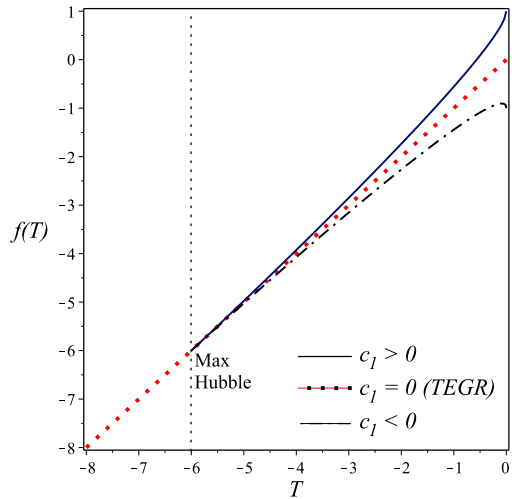}
\caption{\textit{The evolution of the $f(T)$ gravity, Eq. (\ref{Mod1-f(T)}). For $c_1\neq 0$, $f(T)$ cannot be extended beyond $H_{max}$, unlike the TEGR case ($c_1=0$) which has no upper limit on $H$. This is in agreement with the phase portrait pattern of Fig. \ref{Fig:phase-portrait}. We take $M=1$ and $M_{p}=1$.}}
\label{Fig:f(T)}
\end{center}
\end{figure}
It is clear that the GR (or TEGR) limit is recovered by setting $c_{1}=0$. It
is interesting to note that there is no $\beta$-dependence in the
resulting $f(T)$ gravity. Also, as it can be seen from Fig.
\ref{Fig:f(T)}, for $c_1= 0$, the TEGR theory has no upper limit on $H$ which is compatible with the Big-Bang scenario. However, for the $c_1 \neq 0$ cases the $f(T)$ theory cannot be extended beyond $H_{max}$, where the initial singularity becomes of Type II or IV according to the sign of $c_1$. In both cases there are an upper limit on $H$. This is in agreement with the phase portraits given in Fig. \ref{Fig:phase-portrait}. Also, at late time for non-zero values of $c_1$ the phase portrait does not evolve towards Minkowski as in the TEGR case, but instead it evolves towards de Sitter or towards a collapsing phase. In Fig. \ref{Fig:f(T)}, we exaggerate the effect of the constant $c_1$ to display the deviation from TEGR clearly. However, in physical scenarios we expect the value of $c_1$ to be tiny enough to be compatible with TEGR at the intermediate phase. This is necessary to provide the thermal history matching the standard cosmology \cite{Awad:2017yod}. Although the model at hand shows many interesting qualitative behaviours, it needs more investigation to examine its viability on the quantitative level.
\subsection{Examining the WEC of the Model}\label{Sec:5.2}
It has been shown earlier in Sec. \ref{Sec:4} that the WEC in the case of a canonical scalar field minimally coupled to $f(T)$ gravity reduces to $f-Hf'\geq 0$ and $\dot{H}f'' \geq 0$. In this section, we examine the validity of the WEC for the model at hand, which provides a good tool to constrain the viable range of the constant-roll parameter $\beta$. Using (\ref{Mod1-f(H)}) and (\ref{Mod1-phase-portrait1}), we examine the WEC, where
\begin{eqnarray}
\nonumber \rho_{\phi}\geq 0&\Rightarrow& f-Hf'=\frac{2}{M}\left(3MH^2+c_1 M_p \sqrt{M^2-H^2}\right)\geq 0, \\
\nonumber  \rho_{\phi}+p_{\phi}=\dot{\phi}^2\geq 0 &\Rightarrow& \dot{H}f''=-\frac{4\beta}{M}\left(3 M H^2+c_1 M_p \sqrt{M^2-H^2}\right)\geq 0.
\end{eqnarray}
\begin{figure}[h]
\centering
\subfigure[~$-\frac{3MH^2}{M_p\sqrt{M^2-H^2}}<c_1<0$]{\label{fig:Mod1-WEC1}\includegraphics[scale=.35]{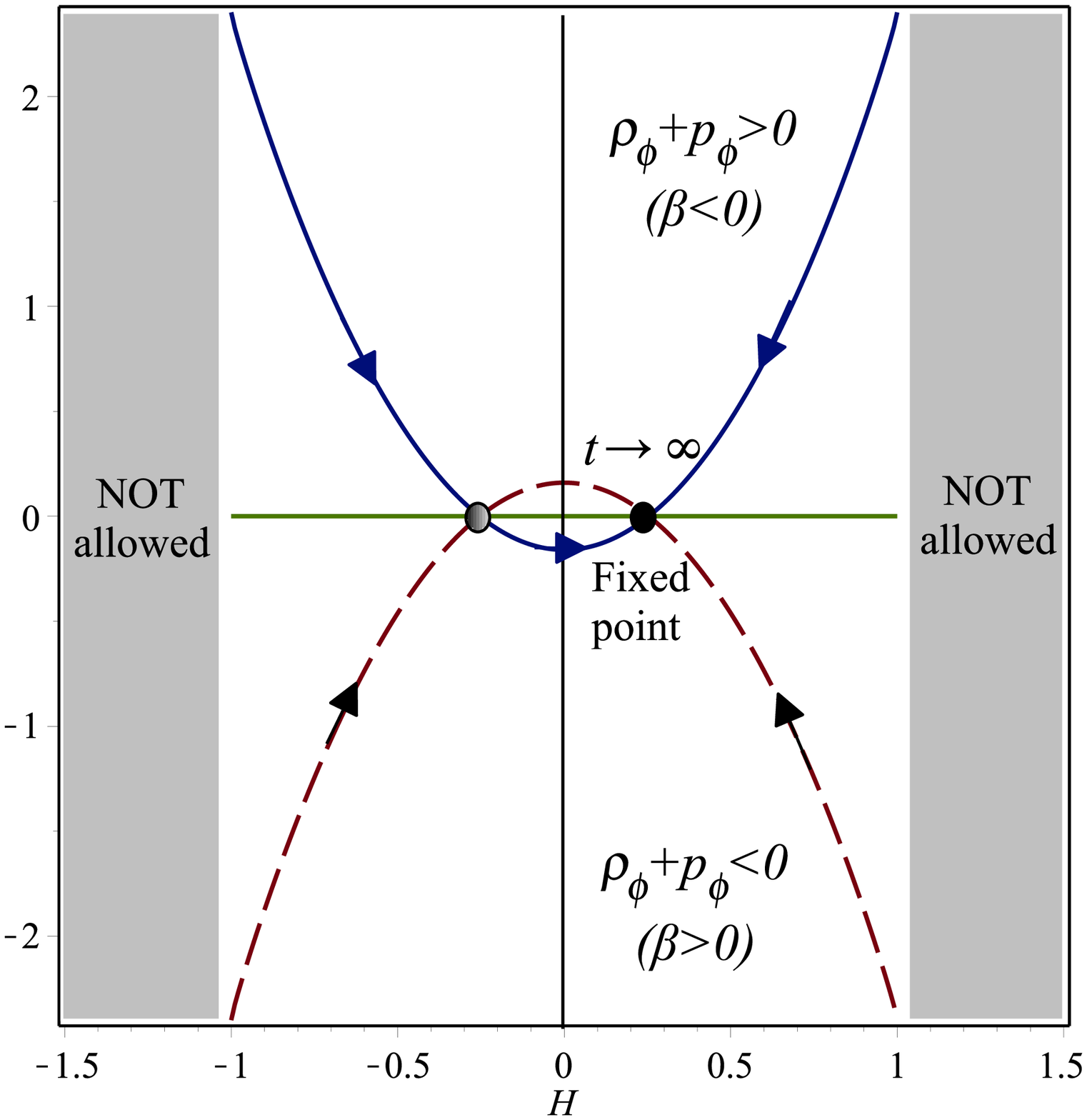}}\hspace{1.5cm}
\subfigure[~$c_1>0$]{\label{fig:Mod1-WEC2}\includegraphics[scale=.35]{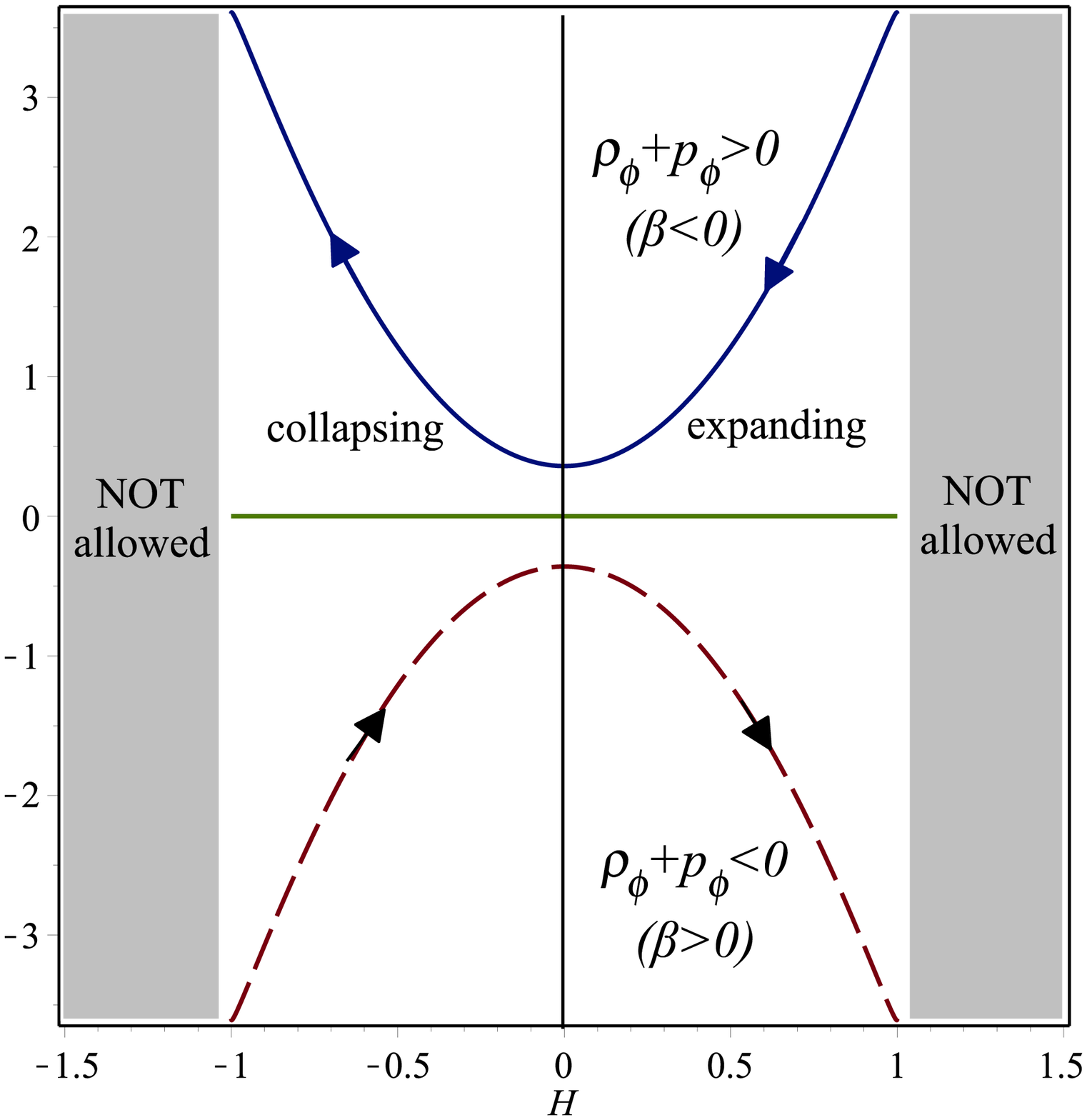}}
\caption[figtopcap]{\textit{Schematic plot of the WEC, $\rho_\phi+p_\phi$, of the $f(T)$ gravity given by (\ref{Mod1-f(T)})}: \subref{fig:Mod1-WEC1} \textit{For $-\frac{3MH^2}{M_p\sqrt{M^2-H^2}}<c_1<0$ and $\beta\leq 0$, the WEC is always fulfilled where the fixed point at small $H$ cannot be reached in a finite time};
\subref{fig:Mod1-WEC2}\textit{ For $c_1>0$ and $\beta\leq 0$, WEC is always fulfilled, while transition from expansion $H>0$ to contraction $H<0$ can be realized by the model. We take $M=1$ and $M_p=1$}.}
\label{Fig:Mod1-WEC}
\end{figure}
The solution of the above system of inequalities implies that $c_1\geq -\frac{3MH^2}{M_p\sqrt{M^2-H^2}}$ and $\beta\leq 0$. We present some plots in Fig. \ref{Fig:Mod1-WEC} that are visualizing the WEC. In Fig. \ref{Fig:Mod1-WEC}\subref{fig:Mod1-WEC1}, we investigate only the possible negative values of the constant $c_1$. It is clear that the WEC is fulfilled as long as $\beta\leq 0$ at the $H>H_{lower-fix}$ patch. We note that $\rho_\phi+p_\phi$ cannot be negative in a finite time, since its null value meets the fixed point $H_{lower-fix}$ as indicated in the Figure. This is in agreement with the phase portraits in Fig. \ref{Fig:Mod1-potentials}\subref{fig:Mod1-pot1} where the viable physical scenario matches $\beta\leq 0$ case. Also, in Fig. \ref{Fig:Mod1-WEC}\subref{fig:Mod1-WEC2}, the WEC is fulfilled as long as $\beta\leq 0$, which is in agreement with the phase portraits given in Fig. \ref{Fig:Mod1-potentials}\subref{fig:Mod1-pot2}. Here we confirm our previous conclusion that the physical scenarios are consistent with the constraint $\beta \leq 0$.
\section{Reconstruction of Constant-roll Inflationary Potentials for a Given $f(T)$ Gravity}\label{Sec:6}

In this section, we shall fix the functional form of the $f(T)$
gravity and we shall investigate which constant-roll potentials does
the $f(T)$ gravity generates. We consider the power-law $f(T)$
gravity of the form,
\begin{equation}\label{power-law-f(T)}
    f(T)=T_{0}\left(\frac{T}{T_{0}}\right)^{n}.
\end{equation}
Substituting Eq. (\ref{power-law-f(T)}) in Eqs. (\ref{generatingH2})
and (\ref{const-roll pot}), we obtain,
\begin{equation}\label{Mod2-Hphi}
    H(\phi)=H_{0}\left[\frac{\beta (n-1)^{2}(\phi-\phi_{0})^{2}}{2n^2 M_{p}^2 (1-2n)}\right]^{\frac{1}{2(n-1)}},
\end{equation}
and also,
\begin{equation}\label{Mod2-pot}
    V(\phi)=V_{0}+(2n-1)(3+\beta)H_{0}^{2}M_{p}^{2}\left[\frac{\beta (n-1)^2 (\phi-\phi_{0})^2}{2n^2 M_p^2 (1-2n)}\right]^{\frac{n}{n-1}}.
\end{equation}
It is clear that $V(\phi)=V_{0}=constant$ and for simplicity, we
take $V=0$ at $\phi=\phi_{0}$, and hence $V_{0}=0$ and $\phi_{0}=0$.
However, by substituting from (\ref{power-law-f(T)}) in the
constant-roll differential equation (\ref{Mod1-phase-portrait2}), we
get,
\begin{equation}\label{Mod2-Hubblet}
    H(t)=-\frac{n}{\beta t- n t_{i}}, \quad a(t)=a_{i}(\beta t - n t_{i})^{-\frac{n}{\beta}},
\end{equation}
where $a_{i}$ and $t_{i}$ are integration constants. One can use Eqs. (\ref{Mod2-Hphi}) and (\ref{Mod2-Hubblet}) to determine $\phi(t)$ explicitly. The above equation characterizes the power-law inflation (PLI), since $a(t)>t^{-n/\beta}$ for $n>0$ and $\beta<0$. Notably, the solutions (\ref{Mod2-Hphi}) and (\ref{Mod2-pot}) are obtained for $n\neq 1$, since $H\propto \phi^{1/(n-1)}$ and $V\propto \phi^{2n/(n-1)}$. So the TEGR ($n=1$) and the non-TEGR ($n\neq 1$) cases should be treated separately in the power-law $f(T)$ gravity. This is similar to the solution of the field equations according to the choice of the EoS $w=-1$ or $w\neq -1$, since the two cases have no smooth transitions. However, if one aims to work within the TEGR limit, he should assume $n=1$ from beginning in Eq. (\ref{power-law-f(T)}). Following the same procedure, he would obtain akin to the standard PLI \cite{Liddle:1988tb} whose $H\propto e^{-\sqrt{-\beta/2}\frac{\phi}{M_p}}$ and $V(\phi)=M_p^2 (3+\beta) e^{-\sqrt{-2\beta}\frac{\phi}{M_p}}$, where the positivity of this potential requires that $-3<\beta<0$. This shows that the TEGR solution is qualitatively different from (\ref{Mod2-Hphi}) and (\ref{Mod2-Hubblet}). We note that the standard PLI is characterized by constant slopes $n_s-1=n_T=2\beta/(1+\beta)<0$ of the power spectra of primordial scalar and tensor perturbations. However, this class of models is ruled out by Planck observation, since it produces a large tensor-to-scalar ratio $r\sim 0.3$. It is worth mentioning that the case $n=1/2$ gives $f(T)\propto \sqrt{T}$ which does not contribute in the field equations and we exclude this case in the discussion. In the following we show that the modification due to the power-law $f(T)$ helps to make this scenario a viable one.

Substituting from Eq. (\ref{Mod2-Hubblet}) in Eq.
(\ref{Hubble-slow-roll}), we obtain,
\begin{equation}\label{Mod2-Hubble-slow-roll}
    \epsilon_{1}=-\frac{\beta}{n}, \quad \epsilon_{N>1}=0.
\end{equation}
Also, the speed of sound of the scalar perturbation in the context
of $f(T)$ gravity is equal to\footnote{We note that the torsion scalar dependence in Eq. (\ref{speed-of-sound}), as is given in \cite{Cai:2011tc,Rezazadeh:2015dza,Rezazadeh:2017edd}, has been replaced by the Hubble parameter which is more appropriate for our analysis.},
\begin{equation}\label{speed-of-sound}
    c_{s}^{2}= \frac{f_{H}}{H f_{HH}}\, ,
\end{equation}
and for the power-law theory of Eq. (\ref{power-law-f(T)}), it
reads,
\begin{equation}\label{Mod2-speed-of-sound}
    c_{s}=\frac{1}{\sqrt{2n-1}}.
\end{equation}
Thus, the causality condition ($c_{s} \leq 1$) sets the constraint
that $n<\frac{1}{2}$ or $n\geq 1$. However, if both the stability
and the causality conditions are imposed ($0 \leq c_{s} \leq 1$),
the parameter $n$ is constrained as $n\geq 1$.
\subsection{Examining the WEC of the power-law $f(T)$ Model}\label{Sec:6.1}
As mentioned in Sec. \ref{Sec:4} that the WEC, in the case of a canonical scalar field minimally coupled to $f(T)$ gravity, provides a good tool to constrain the viable range of the constant-roll parameter $\beta$. In this section we examine the WEC in the case of the power-law $f(T)$ gravity model. Using the Hubble function (\ref{Mod2-Hubblet}), we write the autonomous equation
\begin{equation}\label{Mod2-phase_portr}
\dot{H}(H)=\frac{\beta}{n}H^2.
\end{equation}

\begin{figure}
\centering
\includegraphics[scale=.35]{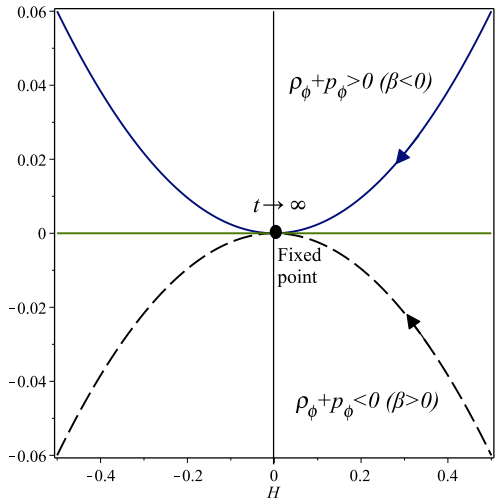}
\caption{\textit{Schematic plot of the WEC, $\rho_\phi+p_\phi$, of the power-law $f(T)$ gravity given by (\ref{power-law-f(T)}). For $n\geq 1$ the WEC is satisfied as long as the constant-roll parameter $\beta<0$. We note that the null value cannot be reached in a finite time, since it meets a fixed point. We take $\beta\neq 0$, $M=1$ and $M_p=1$.}}
\label{Fig:Mod2-WEC}
\end{figure}
The above equation identifies the phase portrait of the model at hand as a parabolic curve (belongs to the standard cosmology model), where its qualitative behaviour shows that the constant-roll $\beta$ should be negative when $n>0$ to match standard cosmology \cite{Awad:2017yod}. Otherwise, the universe evolves towards big-rip in phantom regime. In the case of $\beta=0$, the universe is trapped forever in a fixed point, since this point is de Sitter fixed point. For the $f(T)$ given by (\ref{power-law-f(T)}), the WEC is satisfied as
\begin{eqnarray}
\nonumber \rho_{\phi}\geq 0&\Rightarrow& f-Hf'=6(2n-1)H_0^{2(1-n)}H^{2n}\geq 0, \\
\nonumber  \rho_{\phi}+p_{\phi}=\dot{\phi}^2\geq 0 &\Rightarrow& \dot{H}f''=-12(2n-1)\beta H_0^{2(1-n)}H^{2n}\geq 0.
\end{eqnarray}
Since the sound speed of scalar fluctuations restrict the power-law model to have $n\geq 1$ as shown by Eq. (\ref{Mod2-speed-of-sound}), the constant-roll parameter should follow the constraint $\beta\leq 0$, to satisfy the WEC, confirming the previously obtained results. We also visualize this result graphically as shown in Fig. \ref{Fig:Mod2-WEC}.
\subsection{The Inflationary Parameters of the Power-law $f(T)$ Model}\label{Sec:6.2}
Similar to $k$-inflation models \cite{ArmendarizPicon:1999rj,Garriga:1999vw}, the running of the
speed of sound should be introduced in the $f(T)$ perturbative
analysis, as an additional slow-roll parameter
\cite{Rezazadeh:2015dza,Rezazadeh:2017edd},
\begin{equation}\label{sound-slow-roll}
  s_{1} \equiv -\frac{d \ln c_{s}/dt}{H}, \quad  s_{N+1} \equiv \frac{d \ln s_{N}/dt}{H}.
\end{equation}
We note that the power-law $f(T)$ gravity is characterized by a constant speed of the scalar fluctuations as given by Eq. (\ref{Mod2-speed-of-sound}), subsequently all the sound speed slow-roll parameters (\ref{sound-slow-roll}) of the power law $f(T)$ gravity are null.

In the FLRW cosmological background, small deviations from
homogeneity,
$$\delta\phi(t,\vec{r})=\phi(t,\vec{k})-\phi_{0}(t),$$
can be transformed to Fourier
space, in which case each Fourier mode evolves in an independent way
from the other modes, as it can be seen below,
$$\delta\phi(t,\vec{k})=\int d^{3}\vec{r} e^{-i\vec{k}.\vec{r}}\delta\phi(t,\vec{r}),$$
where $\vec{r}$ and $k=|\vec{k}|$ are the comoving coordinates and
the comoving wavenumber, respectively. Then, $1/k$ defines the
comoving wavelength, and the physical mode wavelength is
$\lambda(t)=a(t)/k$. At sub-horizon scale, the physical wavelength
satisfies $\lambda \ll \lambda_{H}$, where $\lambda_{H}=H^{-1}$ is
the Hubble radius, to which we refer to as ``the horizon''. However,
in $f(T)$ gravity, due to the contribution of the speed of sound of
the scalar fluctuations, it is convenient to modify this condition
to be $\lambda \ll \lambda_{s}$, where $\lambda_{s}=c_{s} H^{-1}$
is the sound horizon.

In observational cosmology, it is convenient to expand the power
spectrum of scalar (tensor) perturbations as follows,
\begin{eqnarray}
  \mathcal{P}_{s}(k) &=& A_{s}\left(\frac{k}{k_{*}}\right)^{n_{s}-1+\frac{1}{2}\frac{dn_{s}}{d\ln k}\ln \left(\frac{k}{k_{*}}\right)+\frac{1}{6}\frac{d^{2}n_{s}}{d \ln k^{2}}\left(\ln \left(\frac{k}{k_{*}}\right)\right)^{2}+\cdots}, \label{scalar-power-spectrum}\\
  \mathcal{P}_{T}(k) &=& A_{T}\left(\frac{k}{k_{*}}\right)^{n_{T}+\frac{1}{2}\frac{dn_{T}}{d\ln k}\ln \left(\frac{k}{k_{*}}\right)+\cdots},\label{tensor-power-spectrum}
\end{eqnarray}
where the wavenumber $k_{*}$ is be set as the pivot scale in the observable range, $A_{s}$ ($A_{T}$) is the scalar (tensor) amplitude,
$n_{s}$ ($n_{T}$), $\frac{d n_{s}}{d \ln k}$ $\left(\frac{d n_{T}}{d
\ln k}\right)$ and $\frac{d^2 n_{s}}{d \ln k^2}$ are the scalar
(tensor) spectral index of primordial curvature perturbations, the
running of the scalar (tensor) spectral index, and the running of
the running of the scalar spectral index, respectively. Following the standard steps of quantization of the theory, we go directly to the spectral density
\begin{equation}\label{Ps}
    \mathcal{P}_s^{\zeta}(k)=\frac{k^3}{2\pi^2}|\zeta_k|^2 =\frac{k^3}{2\pi^2}\frac{|v_k|^2}{z_s^2},
\end{equation}
where $z$ is defined as $z_s=a\sqrt{2\epsilon_1}/c_s$, $\zeta_k$ is a gauge-invariant variable, which characterizes the cosmological inhomogeneities, defined as in \cite{Cai:2011tc}, and the canonical variable $v_k=z_s \zeta_k$. As clear from the above equation that the scale invariant power spectrum requires the leading term to be $|\zeta_k|\propto k^{-3/2}$. Following the procedure given in Ref. \cite{Rezazadeh:2015dza,Rezazadeh:2017edd}, the equation of motion of the scalar fluctuation is written in terms of the canonical variable $v_k$ as
\begin{equation}\label{MS_eqn}
    v''_k+\left(c_s^2 k^2-\frac{z''_s}{z_s}\right)v_k=0,
\end{equation}
where the primes denote derivatives with respect to the conformal time $\tau=\int dt/a$. The above equation reduces to the well-known Mukhanov-Sasaki equation when $c_s=1$. In order to find the scalar power spectrum of the gauge variable $\zeta_k$, we solve (\ref{MS_eqn}), where the initial condition takes the form of Bunch-Davies vacuum
\begin{equation}\label{min_wl}
    v_k \approx \frac{e^{-ikc_s\tau}}{\sqrt{2c_sk}}, \qquad (\lambda \ll \lambda_s).
\end{equation}
Notably, the initial condition above has been modified to include the speed of sound of the scalar fluctuation which has a crucial role in the $f(T)$ gravity on the perturbation level. On the other hand, the long wavelength solution gives
 \begin{equation}\label{long_wl}
    v_k \approx C_k z_s, \qquad (\lambda \gg \lambda_s),
 \end{equation}
where the constant can be fixed by matching the two solutions at the sound horizon crossing ($\lambda=\lambda_s$) in the standard way. In the transition we assume that the universe is quasi-de Sitter $a \sim 1/(H\tau)$ and (\ref{MS_eqn}) becomes a Bessel equation. Along with the initial condition (\ref{min_wl}), it gives \cite{Garriga:1999vw}
\begin{equation}\label{Ck}
    |C_k|^2=\frac{1}{2c_skz_s^2}.
\end{equation}
Thus, at the sound horizon crossing $\lambda=\lambda_s$, i.e. $a=k c_s/H$, we identify the gauge variable $\zeta_k=v_k/z_s$ as below
\begin{equation}\label{zeta_k}
    \left|\zeta_k\right|_{\lambda=\lambda_s}^2=\frac{1}{4M_p^2}\frac{H^2}{c_s k^3 \epsilon_1}.
\end{equation}
It remains now to identify the scalar power spectrum of the gauge variable $\zeta_k$. This can be done by plugging (\ref{zeta_k}) into (\ref{Ps}). This gives the primordial power spectrum of the
primordial curvature scalar perturbations \cite{Cai:2011tc,Rezazadeh:2015dza,Rezazadeh:2017edd}
\begin{equation}\label{Mod2-scalar-power-spectrum}
    \mathcal{P}_{s}(t)=\frac{1}{8\pi^{2}M_{p}^{2}}\left.\frac{H^{2}}{c_{s}\epsilon_{1}}\right|_{\lambda=\lambda_{s}}
    =-\left.\frac{1}{8\pi^{2}M_{p}^{2}}\frac{n^{3}\sqrt{2n-1}}{\beta (\beta t- n t_{i})^{2}}\right|_{\lambda=\lambda_{s}},
\end{equation}
The last equality has been obtained by using (\ref{Mod2-Hubblet}), (\ref{Mod2-Hubble-slow-roll}) and (\ref{Mod2-speed-of-sound}). The sound horizon crossing can be expressed as
$\lambda=\lambda_{s}$, which determines the time of the sound
horizon exit,
\begin{equation}\label{Mod2-sound-hor-exit}
    t_{s}=\frac{n}{\beta k}\left(k t_{i}-\left[\frac{(2n-1)^{1/2} a_{i} k}{n}\right]^{\frac{\beta}{n+\beta}}\right).
\end{equation}
Inserting the above expression in Eq.
(\ref{Mod2-scalar-power-spectrum}), we can evaluate the scalar power
spectrum in terms of the comoving wavenumber $k$, and the resulting
expression is,
\begin{equation}\label{Mod2-scalar-PS-exit}
    \mathcal{P}_{s}(k)=A_s\left(\frac{k}{k_{*}}\right)^{\frac{2\beta}{n+\beta}},
\end{equation}
where
\begin{equation}\label{As}
    A_s=-\frac{n\left(n^2\sqrt{2n-1}~\right)^{\frac{n-\beta}{n+\beta}}}{8\pi^{2}M_{p}^{2}\beta} \left(\frac{nk_{*}}{a_i}\right)^{\frac{2\beta}{n+\beta}}.
\end{equation}

By comparing with (\ref{scalar-power-spectrum}), we easily obtain
the resulting expression for the spectral index of primordial
curvature perturbations, which is,
\begin{equation}\label{Mod2-ns}
    n_{s}-1=\frac{2\beta}{n+\beta}
\end{equation}
Remarkably, the above relation gives a modified version of the
general relativistic power spectrum. We restrict ourselves by
choosing $n_{s}=0.96$ to fulfill the observational constraints from
the joint analysis of Planck and BICEP2/Keck Array collaborations.
Consequently, we get,
\begin{equation}\label{Mod2-beta}
    \beta \simeq -0.02 n.
\end{equation}
The tensor fluctuations power spectrum in $f(T)$ gravity, is given
by the standard expression,
\begin{equation}\label{Mod2-tensor-power-spectrum}
    \mathcal{P}_{T}(k)=\frac{2}{\pi^{2}M_{p}^{2}}\left.\frac{a^{2}H^{2}}{z_{T}^{2}}\right|_{\lambda=\lambda_{H}},
\end{equation}
where the parameter $z_{T}$ is equal to
\cite{Rezazadeh:2015dza,Rezazadeh:2017edd},
\begin{equation}\label{zt}
    z_{T}=a e^{\textstyle{\int}\frac{\gamma}{2}dt},\qquad \gamma=\frac{\dot{T}f_{TT}}{f_{T}}.
\end{equation}
All quantities in the right hand side of Eq.
(\ref{Mod2-tensor-power-spectrum}), should evaluated at the horizon
crossing $\lambda=\lambda_{H}$, where $\lambda_{H}=H^{-1}$ is the
Hubble horizon. In the general case, the freezing out moment of the
scalar fluctuations is determined at the sound horizon crossing time
instance, which is different from the freezing out of the tensor
fluctuations which occur when the Hubble horizon crossing occurs.
However, this difference is negligible if we restrict ourselves to
the lowest order slow-roll parameters \cite{Garriga:1999vw}. In Ref.
\cite{Rezazadeh:2015dza}, it has been proposed that if the following
holds true,
\begin{equation}\label{Non-Gaussianity}
\delta\equiv \frac{|\gamma|}{2H}\ll 1,
\end{equation}
the tensor fluctuations power spectrum in the $f(T)$ gravity reduces
to the standard inflationary model where $z_{T}\approx a$. In order
to check the validity of this condition in the present model, one
may rewrite the parameter $\delta$ as follows,
\begin{equation}\label{Mod2-delta}
    \delta=\frac{\epsilon_{1}}{2}|1-c_{s}^{-2}|.
\end{equation}
Using (\ref{Mod2-Hubble-slow-roll}) and (\ref{Mod2-speed-of-sound}),
$\delta$ becomes,
\begin{equation}\label{Mod2-delta}
    \delta=\left|\frac{\beta(n-1)}{n}\right|=0.02|n-1|,
\end{equation}
where the last quantity in the above equation is evaluated by using
(\ref{Mod2-beta}). We note that the parameter $\delta=0$ in the
TEGR limit ($n=1$), while in the power-law $f(T)$ gravity $\delta < O(1)$ when $-49<n<51$.
Therefore, we find that the condition $\delta \ll 1$ is valid in our
case, and thus, the  power spectrum of the tensor fluctuations matches the standard cosmology case, where
\begin{equation}\label{canonical-tensor-power-spectrum}
    \mathcal{P}_{T}(t)=\left.\frac{2H^{2}}{\pi^{2}M_{p}^{2}}\right|_{\lambda=\lambda_{H}}
    =\left.\frac{2}{\pi^{2}M_{p}^{2}}(\beta t - n t_{i})^{-2}n^{2}\right|_{\lambda=\lambda_{H}},
\end{equation}
We note here that the tensor power spectrum should be estimated at the time of horizon crossing $\lambda=\lambda_H$. This time is not exactly the same as the time of sound horizon crossing $\lambda = \lambda_s$, but to lowest order in the slow-roll parameters this difference is negligible \cite{Garriga:1999vw}. At the horizon exit $\lambda=\lambda_{H}$, we determine the time of the horizon exit,
\begin{equation}\label{Mod2-tensor-horizon-exit}
    t_{T}=\frac{1}{\beta}\left(n t_{i}+(-1)^{\frac{\beta}{n+\beta}}\left[\frac{a_{i} n}{k}\right]^{\frac{\beta}{n+\beta}}\right).
\end{equation}
Inserting (\ref{Mod2-tensor-horizon-exit}) in
(\ref{canonical-tensor-power-spectrum}), we write the tensor power
spectrum in terms of the comoving wavenumber $k$ as follows,
\begin{equation}\label{Mpd2-Pt}
    \mathcal{P}_{T}(k)=A_T\left(\frac{k}{k_{*}}\right)^{\frac{2\beta}{n+\beta}},
\end{equation}
where
\begin{equation}\label{At}
    A_T=\frac{2(-1)^{\frac{2\beta}{n+\beta}}\left(nk_{*}/a_i\right)^{\frac{2\beta}{n+\beta}}}{\pi^{2}M_{p}^{2}}.
\end{equation}

By comparing the above expression, with
(\ref{tensor-power-spectrum}), we get the spectral index of the
tensor power spectrum,
\begin{equation}\label{Mod2-nt}
    n_{T}=\frac{2\beta}{n+\beta}.
\end{equation}
Interestingly, Eqs. (\ref{Mod2-ns}) and (\ref{Mod2-nt}) identify that the slopes of the power spectra $n_s-1=n_T=\frac{2\beta}{n+\beta}$ are constant similar to the standard PLI except the newly introduced $n$ parameter. Using (\ref{Mod2-beta}), we find that $n_{T}\simeq -0.04$. It is worth to mention that this value does not depend on the index $n$.
Thus, the scale-dependance of the tensor fluctuations power spectrum
(\ref{tensor-power-spectrum}) can be measured by using the spectral
index,
\begin{equation}
    n_{T}\equiv \frac{d\ln \mathcal{P}_{T}}{d\ln k}=-2\epsilon_{1}.\label{tensor-indx}
\end{equation}
This observable is not measured accurately up to date, however,
using (\ref{Mod2-scalar-power-spectrum}) and
(\ref{canonical-tensor-power-spectrum}), the scalar-to-tensor ratio
in $f(T)$ is given by,
\begin{equation}\label{tensor-to-scalar}
    r\equiv \frac{\mathcal{P}_{T}}{\mathcal{P}_{s}}=16c_{s}\epsilon_{1}=-8c_{s}n_{T}.
\end{equation}
Remarkably, there is no way to put an upper limit on the parameter
$\epsilon_{1}$ from the above relation, without constraining the
speed of sound. However, it reduces to the standard consistency
relation by setting $c_{s}=1$. From (\ref{Mod2-speed-of-sound}) and
(\ref{Mod2-nt}), we find,
\begin{equation}\label{Mod2-r}
    r=-\frac{16\beta}{(n+\beta)\sqrt{2n-1}}=\frac{0.32}{\sqrt{2n-1}},
\end{equation}
where the last quantity in the above equation is evaluated by using
(\ref{Mod2-beta}). Although the modified PLI is qualitatively different from the standard one, it leads to constant slopes,
$$n_s-1=n_T=\frac{2\beta}{(n+\beta)}<0,$$
of the power spectra of primordial scalar and tensor perturbations, as obviously seen from (\ref{Mod2-ns}) and (\ref{tensor-indx}), just as in the standard PLI. Interestingly, for $n\gtrsim 4.73$, the model fulfills the upper bound of the Planck data  $r_{0.002}<0.11$ at 95$\%$ CL. This is on the contrary to the excluded standard PLI ($n=1$) which produces a large $r = 0.32$. Thus, we conclude that the power-law $f(T)$ gravity provides a better frame work with that scenario. Finally, we summarize some numerical values of the model parameters, for different choices of the parameter $n$ in Table \ref{Table:Mod2-parameters}.
\begin{table}[h]
  \centering
  \caption{Constant-roll inflation of the power-law $f(T)$ gravity}\label{Table:Mod2-parameters}
  \begin{tabular}{|c|c|c|c|c|c|c|}
    \hline\hline
    $n$ & $\beta$ & $c_{s}$ & $n_{s}$ & $\delta$ & $n_{T}$ & $r$ \\
    \hline
    $4.73$   & $-0.093$ & $0.0.344$  & $0.96$  & $0.073$  & $-0.04$  & $0.11$  \\
    $5$     & $-0.098$ & $0.333$  & $0.96$  & $0.078$  & $-0.04$  & $0.10$  \\
    $6$     & $-0.118$ & $0.301$  & $0.96$  & $0.098$  & $-0.04$  & $0.09$ \\
    $7$     & $-0.137$  & $0.277$  & $0.96$  & $0.118$  & $-0.04$  & $0.08$  \\
    \hline
    Eq. No.&(\ref{Mod2-beta})&(\ref{Mod2-speed-of-sound})&(\ref{Mod2-ns})&(\ref{Mod2-delta})&(\ref{Mod2-nt})&(\ref{Mod2-r})\\
    \hline\hline
  \end{tabular}
\end{table}
As it can be seen in Table \ref{Table:Mod2-parameters}, the compatibility with the Planck data occurs for a wide range of the
free parameters of the model. In conclusion, within the power-law $f(T)$ gravity, the constant-roll parameter is slightly less than zero, particularly $\beta \lesssim -0.093$, as shown in Table \ref{Table:Mod2-parameters}. Thus, according to Eq. (\ref{Mod1-beta}), we expect the inflaton to have an EoS slightly above $-1$. So we should not worry about ghost instability problems in our case. Notably, when the constant-roll potential (\ref{pot2}) includes a non-vanishing $V_0$-term, we could obtain positive values of $\beta$, while the WEC is still valid. This case is similar to the constant-roll model which has been obtained in Ref. \cite{Motohashi:2014ppa,Motohashi:2017aob}.
\section{Concluding Remarks}\label{Sec:7}
In this paper we investigated the implications of a constant-roll
condition on $f(T)$ gravity inflation. We assumed that the theory is
described by an inflaton minimally coupled to an $f(T)$ teleparallel
gravity, and we examined in detail the implications of the
constant-roll condition in the cosmological evolution. Our approach
enabled us to introduce a reconstruction technique, in the context
of which it is possible by fixing the Hubble evolution, to find both
the constant-roll scalar potential and also the $f(T)$ gravity which
may generate such evolution. Also, by fixing the $f(T)$ gravity, by
employing the reconstruction technique we developed, we were able to
find both the Hubble rate corresponding to it and also the scalar
potential. Also we calculated the power spectrum of primordial
scalar curvature perturbations and also the power spectrum of
primordial tensor perturbations, and we investigated the
implications of the constant-roll condition on the spectral index
and the scalar-to-tensor ratio. As we showed, the resulting
observational indices can be compatible with the observational data,
and we examined the parameter space in order to find which values
allow the compatibility with current observational data.

As a general conclusion, by taking into account the results of the
present study but also of previous studies of $F(R)$ gravity, the
constant-roll condition can provide an appealing theoretical
framework, in the context of which a viable theory of inflation is
obtained, which is compatible with the current observational data.
What now remains, is to investigate the implications of the
constant-roll scenario on Gauss-Bonnet $F(G)$ theories, and also
other modified gravity theories such as mimetic gravity or $F(R,T)$
gravity.
\acknowledgments
The authors would like to thank the anonymous referee for her/his careful reading and suggestions which indeed helped to improve the manuscript. This work is supported by the Egyptian Ministry of Scientific Research under Project No. 24-2-12 (A.A, W.E and G.G.L.N), (MINECO, Spain) project FIS2016-76363-P and by (AGAUR, Catalonia) project 2017 SGR247 (S.D.O.).

\providecommand{\href}[2]{#2}\begingroup\raggedright\endgroup
\end{document}